\documentclass[sigconf,screen]{acmart}
\usepackage[utf8]{inputenc}
\usepackage[table]{xcolor} 
\usepackage{pifont}
\usepackage{menukeys} 
\usepackage{paralist}
\usepackage{amsmath}
\usepackage{multirow}
\usepackage{epsfig,endnotes}
\usepackage{tikz}
\usepackage[ruled,vlined,linesnumbered]{algorithm2e}
\usepackage{amsfonts}
\usepackage{url}
\usepackage{booktabs}
\usepackage{multicol}
\usepackage{lipsum}
\usepackage{adjustbox}
\usepackage{tabularx}
\usepackage[english]{babel}
\usepackage{listings}
\usepackage{subcaption}
\usepackage{graphicx}
\usepackage{wrapfig}
\usepackage{tcolorbox}
\usepackage{enumitem}
\usepackage{circledsteps}
\usepackage{makecell}
\usepackage{arydshln}
\usepackage{threeparttable}
\usepackage{hyperref}
\usepackage{textcase}
\usepackage{color,colortbl}
\usepackage{array}
\usepackage{nicematrix}
\usepackage{balance}

\definecolor{myblue}{HTML}{3c7fb1}
\definecolor{myyellow}{HTML}{f5a65b}
\definecolor{mygreen}{HTML}{44a05c}

\definecolor{customred}{HTML}{FF5314}
\definecolor{customgreen}{HTML}{7CBB00}

\AtBeginDocument{%
  }

\newcolumntype{C}[1]{>{\centering\arraybackslash}p{#1}}

\copyrightyear{2026}
\acmYear{2026}
\setcopyright{cc}
\setcctype{by}

\acmConference[ASE '26]{Proceedings of the 41st IEEE/ACM International Conference on Automated Software Engineering}{October 12--16, 2026}{Munich, Germany}
\acmBooktitle{Proceedings of the 41st IEEE/ACM International Conference on Automated Software Engineering (ASE '26), October 12--16, 2026, Munich, Germany}
\acmDOI{10.1145/3832783.3834339}
\acmISBN{979-8-4007-2882-2/2026/10}
\acmSubmissionID{ase26main-p130-p}
\received{2026-03-26}
\received[accepted]{2026-06-18}

\author{Xiuwei Shang}
\authornote{Xiuwei Shang is also affiliated with Singapore Management University, Singapore.}
\orcid{0009-0009-6660-9947}
\email{shangxw@mail.ustc.edu.cn}
\affiliation{%
  \institution{University of Science and Technology of China}
  \city{Hefei}
  \country{China}
}

\author{Li Hu}
\orcid{0009-0001-7857-7454}
\email{pdxbshx@mail.ustc.edu.cn}
\affiliation{%
  \institution{University of Science and Technology of China}
  \city{Hefei}
  \country{China}
}

\author{Xiao Jiang}
\orcid{0009-0002-6739-6102}
\email{jx\_xiao@mail.ustc.edu.cn}
\affiliation{%
  \institution{University of Science and Technology of China}
  \city{Hefei}
  \country{China}
}

\author{Jieke Shi}
\authornote{Jieke Shi and Shaoyin Cheng are the corresponding authors.}
\orcid{0000-0002-0799-5018}
\email{jiekeshi@smu.edu.sg}
\affiliation{%
  \institution{Singapore Management University}
  \country{Singapore}
}

\author{Junda He}
\orcid{0000-0003-3370-8585}
\email{jundahe.2022@phdcs.smu.edu.sg}
\affiliation{%
  \institution{Singapore Management University}
  \country{Singapore}
}

\author{Zhou Yang}
\orcid{0000-0001-5938-1918}
\email{zy25@ualberta.ca}
\affiliation{%
  \institution{University of Alberta \& Alberta Machine Intelligence Institute}
  \city{Edmonton}
  \country{Canada}
}

\author{Shaoyin Cheng}
\authornotemark[2]
\orcid{0000-0002-3992-9509}
\authornote{Also affiliated with Anhui Province Key Laboratory of Digital Security, Hefei, China.}
\email{sycheng@ustc.edu.cn}
\affiliation{%
  \institution{University of Science and Technology of China}
  \city{Hefei}
  \country{China}
}

\author{Guoqiang Chen}
\orcid{0000-0003-0651-6617}
\email{ch3nye@mail.ustc.edu.cn}
\affiliation{%
  \institution{University of Science and Technology of China}
  \city{Hefei}
  \country{China}
}

\author{Weiming Zhang}
\authornotemark[3]
\orcid{0000-0001-5576-6108}
\email{zhangwm@ustc.edu.cn}
\affiliation{%
  \institution{University of Science and Technology of China}
  \city{Hefei}
  \country{China}
}

\author{David Lo}
\orcid{0000-0002-4367-7201}
\email{davidlo@smu.edu.sg}
\affiliation{%
  \institution{Singapore Management University}
  \country{Singapore}
}

\renewcommand{\shortauthors}{Xiuwei Shang et al.}

\newcommand{\bench}{{\sc BinJudgeBench}\xspace}
\newcommand{\method}{{\sc BinJudge}\xspace}

\begin{document}

\title{Beyond Text Matching: Towards Reference-Free Evaluation for Human-Oriented Binary Reverse Engineering}

\begin{abstract}
Human-Oriented Binary Reverse Engineering (HOBRE) aims to transform decompiled pseudocode into a more human-friendly representation, thereby reducing the cognitive burden of reverse analysis and improving efficiency. However, reliably evaluating HOBRE outputs remains a fundamental challenge. On one hand, human evaluation is costly, time-consuming, as well as difficult to scale. On the other hand, existing automated metrics either rely on executable test cases and runtime environments, which are often unavailable for real-world binaries, or depend on high-quality source code references that are typically inaccessible and fail to capture semantically equivalent but lexically diverse outputs. Although the emerging LLM-as-a-Judge paradigm is naturally well-suited to HOBRE evaluation, its effectiveness has not yet been fully studied.

This paper presents the first systematic investigation of the LLM-as-a-Judge paradigm for HOBRE, covering three representative tasks: function name recovery, binary code summarization, and decompilation optimization. We introduce \bench, the first expert-annotated, reference-free evaluation benchmark based on multi-dimensional human judgment. Our empirical study reveals that LLM-as-a-Judge achieves an average correlation of 63.20\% with human judgment, significantly outperforming traditional automated metrics at 35.04\%. By analyzing the impact of various judge configurations, including backbone LLMs, prompting strategies, and decoding temperatures, on both correlation and cost, we find that no ``one-size-fits-all'' configuration exists, as the optimal setup varies across tasks and individual samples. To address this, we propose \method, which employs a lightweight routing mechanism to adaptively select the optimal judge configuration for each specific task and sample. \method improves correlation with human experts by 4.5\%-24.7\% and reduces API cost to 0.06$\times$-0.84$\times$ of that of static best configurations, providing a scalable, cost-effective, and high-fidelity automated evaluation scheme for HOBRE.


\end{abstract}



\maketitle

\vspace{-2.1ex}
\section{Introduction}\label{sec:intro}
\vspace{-1.1ex}
Binary reverse engineering is the cornerstone of security-critical scenarios such as malware analysis~\cite{Hung2023SOICT}, software vulnerability detection~\cite{lu2025malsight}, and patch analysis~\cite{xu2017spain}, where reverse analysts often lack access to source code and must confront obscure machine code directly. Modern decompilers (e.g., IDA Pro's Hex-Rays~\cite{IDA}) have become the industry standard, lifting low-level assembly into C-style pseudocode to lower the barrier to analysis. Nevertheless, due to aggressive compiler optimizations and symbol table stripping, the resulting pseudocode often lacks essential semantic information, such as meaningful function names and developer comments, and suffers from distorted code structures~\cite{li2025pseudofix,zhou2025fidelitygpt}, making reverse engineering still a cognitively demanding and labor-intensive task~\cite{shang2025binmetric}.

Driven by the advancements in neural machine translation and large language models, \textbf{Human-Oriented Binary Reverse Engineering (HOBRE)}~\cite{su2024source,chen2022augmenting} has emerged. Researchers leverage deep learning models to post-process decompiled pseudocode, aiming to elevate it into human-readable content that resembles source code and bridge the semantic gap between binary and high-level source code. Such efforts include recovering meaningful function names~\cite{jin2022symlm}, generating natural language summaries~\cite{xiong2023hext5}, and reconstructing pseudocode to improve readability~\cite{tan2024llm4decompile, tan2025sk2decompile}, thereby significantly reducing the cognitive burden on analysts and empowering downstream expert analysis tasks.

\begin{figure*}[ht]
	\centering
        \scalebox{0.9}{
	\includegraphics[width=1\linewidth]{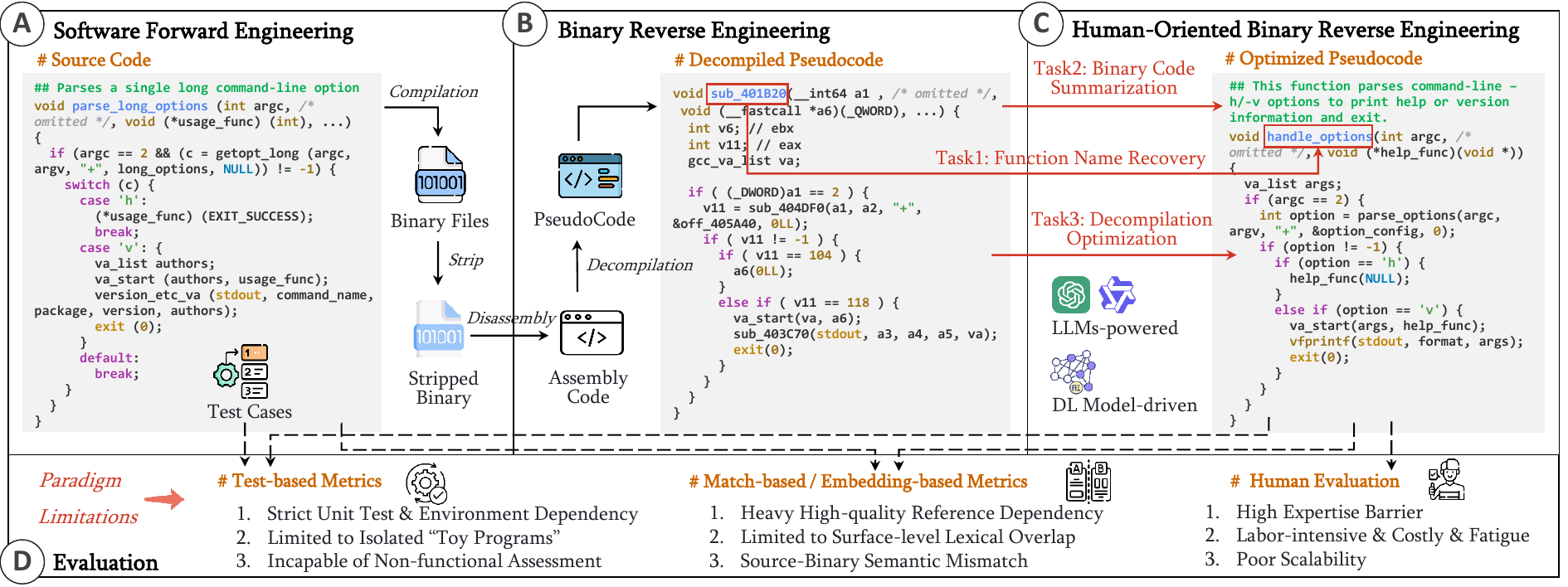}
        }
    \vspace{-2ex}
	\caption{Background of Human-Oriented Binary Reverse Engineering (HOBRE)}
    \vspace{-2ex}
    \label{fig:background}
\end{figure*}

\noindent \underline{\textbf{\emph{Limitations of Existing Metrics.}}} However, effective evaluation of these HOBRE tasks remains significant challenges. First, while \textbf{\emph{human evaluation}} is regarded as the ``gold standard'' for verifying model performance, it has obvious limitations in large-scale assessment. The high barrier to entry in binary reverse engineering makes human auditing not only expensive and time-consuming but also prone to consistency issues, as analysts suffer from fatigue and diminished attention when processing massive datasets. Second, several decompilation optimization studies, exemplified by LLM4Decompile~\cite{tan2024llm4decompile}, employ \textbf{\emph{test-based metrics}} such as re-executability rate~\cite{armengol2022exebench,wong2023refining} to verify functional correctness. These metrics, however, require comprehensive unit tests and execution environments, limiting their application to single-function or single-file ``toy programs''. In practical reverse engineering scenarios, complex binary projects often lack corresponding unit test cases, and their environmental dependencies are extremely difficult to simulate, making dynamic verification infeasible. Furthermore, such metrics are unable to evaluate non-functional aspects like human readability~\cite{weyssowcodeultrafeedback}, nor can they be applied to tasks like function name recovery or binary code summarization.

To achieve scalable evaluation for HOBRE tasks, several traditional automated evaluation metrics have been considered. One type is \textbf{\emph{match-based metrics}} such as BLEU~\cite{papineni2002bleu} and CodeBLEU~\cite{ren2020codebleu}, which assess surface-level overlap between generated content and references (i.e., ground truth). Another type is \textbf{\emph{embedding-based metrics}}, such as BERTScore~\cite{zhang2019bertscore} and CodeBERTScore~\cite{zhou2023codebertscore}, which first encode both generated and reference texts into embeddings and then measure their similarity. However, these metrics also possess inherent limitations. First, they rely on high-quality and accessible reference texts. In HOBRE tasks, references are typically derived from explicit information in the source code (e.g., function names and developer comments). Yet, in the most pressing real-world scenarios beyond experimental environments, such as malware analysis and vulnerability discovery in closed-source firmware, source code is often unavailable, rendering evaluation methods based on source-derived references ineffective. Even in open-source software, only a small fraction of functions contain usable developer comments, which vary significantly in quality and style, making them unreliable as stable evaluation ground truth. Most importantly, after compilation optimization and stripping, the source code exhibits a fundamental semantic mismatch with its corresponding binary code; thus, forcibly using source code content as ground truth for text matching lacks justification. Furthermore, these metrics focus more on lexical similarity than semantic similarity, whereas correct results often exist in various forms that are semantically equivalent but phrased differently.

\vspace{0.1ex}
\noindent \underline{\textbf{\emph{Potential of LLM-as-a-Judge in HOBRE Evaluation.}}}
Recent advancements in LLMs have fostered the rise of the \textbf{\emph{LLM-as-a-Judge}} \cite{li2025generation} paradigm, making it a scalable and cost-effective alternative to human expert evaluation. By guiding LLMs to perform automated audits of generated content based on predefined criteria, this paradigm has demonstrated high human alignment in tasks like code generation~\cite{yang2025code} and program repair~\cite{zhou2025se}, showcasing its significant promise within the software engineering community~\cite{he2026llm}.

Similarly, several key attributes make LLMs particularly suitable for evaluating the HOBRE tasks. First, existing research indicates that LLMs exhibit a certain level of binary program comprehension~\cite{shang2024far,shang2026empirical} and human-like reasoning abilities~\cite{guo2025deepseek}. Second, LLMs are typically trained through reinforcement learning from human feedback (RLHF)~\cite{kaufmann2024survey}, aligning closely with human judgment preferences~\cite{weyssowcodeultrafeedback} and naturally fitting the objectives like "readability" and "usefulness" of HOBRE. Most importantly, unlike metrics like BLEU that mechanically compare text similarity, LLMs excel at handling "semantically equivalent but diversely expressed" scenarios, while leveraging internal reasoning to directly verify generated artifacts against the code's inherent logic, obviating the dependency on explicit reference text, which is critical for "in-the-wild" binary reverse engineering where original source code is inaccessible. 
Nevertheless, the evaluation reliability of LLM-as-a-Judge in the specialized domain of binary reverse engineering, and the degree of its consistency with human experts, has not yet been systematically investigated.

\vspace{0.1ex}
\noindent \underline{\textbf{\emph{Benchmark and Empirical Study.}}} 
To this end, this paper proposes an evaluation perspective that \emph{goes beyond text matching}, systematically investigating for the first time the applicability and capability boundaries of LLM-as-a-Judge in HOBRE tasks, covering three of the most representative tasks: function name recovery, binary code summarization, and decompilation optimization.  

To enable a reliable empirical study, we construct \bench, which is carefully annotated by three human experts in reverse engineering. \bench records human experts' judgments across multiple quality dimensions, providing a curated benchmark for reference-free evaluation. Based on this, we systematically compare LLM-as-a-Judge with traditional automated metrics in terms of their correlation with human expert judgment, and further analyze how different judge configurations, including backbone LLMs, prompting strategies, and decoding temperatures, affect both correlation and cost. Our empirical results reveal that LLM-as-a-Judge substantially outperforms traditional metrics in approximating human judgment, achieving an average correlation of 63.20\% compared to 35.04\% for traditional metrics. Moreover, the optimal judge configuration varies across tasks and individual samples, indicating that no ``one-size-fits-all'' configuration exists.


\vspace{0.2ex}
\noindent \underline{\textbf{\emph{Proposed Approach.}}} 
Motivated by the empirical findings, we further propose \method, a lightweight routing mechanism designed for task- and sample-aware adaptive configuration selection, which employs a distribution-aligned training strategy, explicitly optimizing both evaluation correlation and cost. By dynamically selecting the optimal judge configuration for each sample, \method improves the correlation with human experts by 4.5\%-24.7\% while reducing API cost to only 0.06$\times$-0.84$\times$ of that of static configurations, effectively narrowing the gap between the static best configurations and oracle upper bound. This provides a scalable, cost-effective, and high-fidelity automated evaluation solution for the HOBRE tasks. 


\vspace{0.2ex}
\noindent \textbf{Contributions.} Our major contributions are as follows:
\setlength{\leftmargini}{15pt}
\begin{itemize}
    \item \textbf{\bench Benchmark:} We release the first high-quality, expert-annotated dataset specifically designed for reference-free HOBRE evaluation. It contains 1,233 meticulously audited samples, providing a reliable ground-truth based on multi-dimensional human judgment.
    \item \textbf{Systematic Empirical Study:} We conduct the first comprehensive study to explore the effectiveness of LLM-as-a-judge for HOBRE evaluation, analyzing its correlation with human experts across three core tasks. 
    \item \textbf{\method Router:} We propose a lightweight adaptive routing mechanism that can dynamically select the optimal judge configuration for each individual sample, thereby improving correlation while reducing costs.
\end{itemize}

\vspace{-1ex}
\section{Background and Related Works}\label{sec:back}

\subsection{Human-Oriented Binary Reverse Engineering (HOBRE)}\label{sec:hobre}

As illustrated in Figure \ref{fig:background}, traditional binary reverse engineering primarily focuses on ``recovering program behavior from binaries,'' and its outputs are often intermediate results oriented toward machine semantics, including disassembled instruction sequences, control-flow graphs, data-flow relations, and C-like pseudocode produced by decompilation~\cite{sutherland2006empirical}. Although modern decompilers have made substantial progress in recovering logical equivalence, these outputs remain at a low semantic level. While they can also support downstream tasks such as vulnerability detection and malware analysis, a cognitive gap still exists for human analysts: on the one hand, compiler optimizations can significantly alter the forms of control and data structures~\cite{xu2025novel,xu2025simtam}; on the other hand, symbol stripping removes critical semantic cues like function names, variable names, and type information, making the analysis process heavily reliant on expert experience and extensive manual reasoning.

To bridge the aforementioned gap, HOBRE shifts the perspective toward a goal that is closer to ``human readability and usability''~\cite{su2024source,chen2022augmenting}: \emph{taking decompiled pseudocode as the core input, it performs semantic completion and expression optimization to produce a more human-oriented representation that is closer to source-level expression.} Therefore, in contrast to classical ``Software Forward Engineering'' (source code $\rightarrow$ compilation $\rightarrow$ binary), HOBRE can be regarded as a post-processing paradigm that, building upon ``Traditional Binary Reverse Engineering'' (binary $\rightarrow$ decompilation $\rightarrow$ decompiled pseudocode), takes an additional step toward a more human-friendly high-level representation.

\vspace{-1ex}
\subsection{Target Tasks}\label{sec:relatedworks} 
To comprehensively cover the research scope of HOBRE, this paper focuses on the following three core tasks, which are representative and complementary across semantic dimensions, spanning identifier-level function name recovery, natural-language–level summarization, and program-level decompilation optimization. Together, they cover two major task paradigms, i.e., Code2NL and Code2Code, and encompass three typical output forms, i.e., short-form text, long-form text, and structured code.

\subsubsection{Function Name Recovery (FNR)}  The goal of this task is to rename anonymous functions in stripped binaries with descriptive names (e.g., \texttt{sub\_401B20} → \texttt{handle\_options} in Figure \ref{fig:background}), which provide direct clues to the program's intent. Early research mainly relied on heuristic rules or similarity-based retrieval~\cite{ming2012ibinhunt,he2018debin}. In recent years, deep learning approaches (e.g., NFRE~\cite{gao2021lightweight}, SymLM~\cite{jin2022symlm}, and Llasm~\cite{sha2025llasm}) have become dominant, achieving notable progress by learning mappings between assembly or pseudocode features and source-level symbols. 
The evaluation challenge of this task lies in the inherently highly diverse and style dependence of function naming, as well as the prevalence of domain-specific abbreviations and acronyms, both of which make hard string-matching metrics inadequate for fair assessment.

\renewcommand{\arraystretch}{0.98}
\setlength{\arrayrulewidth}{.5pt}
\setlength{\tabcolsep}{3pt}
\begin{table*}[t]
\centering
\footnotesize
\vspace{-2.3ex}
\caption{Human Annotation Guideline (\textcolor{mygreen}{[green]} \textcolor{myyellow}{[yellow]} \textcolor{myblue}{[blue]} represent task-specific information for Function Name Recovery, Binary Code Summarization, and Decompilation Optimization tasks, respectively.)}
\label{tab:guideline}
\vspace{-1em}
    \begin{tabular*}{\linewidth}{p{.99\linewidth}}
    \toprule
For each instance, annotators are provided with the \textbf{stripped decompiled pseudocode $P_f$} and the \textbf{candidate output $O_f$} generated by the model for the \textcolor{mygreen}{[Function Name Recovery]} \textcolor{myyellow}{[Binary Code Summarization]} \textcolor{myblue}{[Decompilation Optimization]} task (i.e., \textcolor{mygreen}{[Function Name]} \textcolor{myyellow}{[Summary]} \textcolor{myblue}{[Optimized Pseudocode]}). 

The pseudocode’s project origin, corresponding source code $S_f$, and compilation settings are provided for reference, but the model's identity remains anonymous. 

The annotator needs to assign a single overall score (1–5 Likert scale) for each instance, which jointly reflects the following three dimensions:

\vspace{0.2ex}
\textbf{Evaluation Dimensions}: \\
\:\: 1. \textit{Semantic Correctness}: To what extent does the output $O_f$  faithfully reflect the semantics and functional intent of the given pseudocode $P_f$?\\
\:\: 2. \textit{Reverse Analyst Utility}: To what extent does the output $O_f$ reduce the cognitive load for the reverse engineer? Is it closer to a higher-level representation of the source code $S_f$? \\
\:\: 3. \textcolor{mygreen}{[\textit{Naming Distinctiveness \& Naturalness}: Is the output $O_f$  specific and natural, effectively distinguishing between similar functions, and adhering to common engineering naming conventions?]} \\
\:\:\:\: \textcolor{myyellow}{[\textit{Information Coverage \& Density}: Does the output $O_f$ cover the key operations while avoiding lengthy, line-by-line translation and providing high-level abstraction?]} \\
\:\:\:\: \textcolor{myblue}{[\textit{Idiomization \& Faithfulness}: Does the style and structure of the output $O_f$ conform to human programming habits while avoiding the addition of overly speculative semantics?]} \\
\textbf{Scoring}: Rate outputs on a scale of 1 to 5: \\
\:\: \textit{$\cdot$5 (Excellent)}: Highly accurate and faithful, significantly reducing the reverse analysts’ cognitive burden, \textcolor{mygreen}{[name is specific and natural, clearly distinguishing similar functions.]} \textcolor{myyellow}{[summary covers key operations with high information density, avoiding verbose translation.]} \textcolor{myblue}{[optimized pseudocode is highly idiomatic and free of any groundless speculation.]}\\
\:\: \textit{$\cdot$4 (Good)}: Mostly correct and effectively reduces cognitive burden, with only minor, non-misleading deviations, \textcolor{mygreen}{[name is correct but slightly generic, following basic naming conventions.]} \textcolor{myyellow}{[summary covers nearly all key operations but is slightly wordy.]} \textcolor{myblue}{[optimized pseudocode follows most habits with very few groundless guesses.]} \\
\:\: \textit{$\cdot$3 (Acceptable)}: Partially correct but of limited value; helpful to the analyst but contains some deviations that could be misleading, \textcolor{mygreen}{[name is basically correct but too broad, partially violating naming conventions.]}\textcolor{myyellow}{[summary covers basic functions but is quite verbose.]}\textcolor{myblue}{[optimized pseudocode follows basic habits but contains some unsupported speculation.]} \\
\:\: \textit{$\cdot$2 (Poor)}: Contains obvious semantic errors and offers minimal help, with significant misleading information, \textcolor{mygreen}{[name is weakly related and violates naming conventions.]}\textcolor{myyellow}{[summary severely misses operations and contains meaningless descriptions.]}\textcolor{myblue}{[optimized pseudocode is generally non-idiomatic and increases auditing burden due to over-speculation.]} \\
\:\: \textit{$\cdot$1 (Bad)}: Fundamentally incorrect, fails to support analysis, and severely misleads the analyst, \textcolor{mygreen}{[name is entirely unrelated or uses random/meaningless characters.]}\textcolor{myyellow}{[summary is irrelevant, consisting entirely of verbose and meaningless descriptions.]}\textcolor{myblue}{[optimized pseudocode is completely non-idiomatic and consists entirely of model hallucinations.]} \\

    \arrayrulecolor{black}
    \bottomrule
    \end{tabular*}
    \vspace{-2ex}
\end{table*}

\subsubsection{Binary Code Summarization (BCS)}  
This task aims to generate concise natural-language descriptions that explain the functionality of a given function. With the advance of multi-modal pretraining, models such as BinT5~\cite{al2023extending} and HexT5~\cite{xiong2023hext5} attempt to align binary features with natural-language descriptions within a unified representation space. Evaluation also remains challenging because summarization is a typical open-ended generation problem in which a single function can be described in many semantically equivalent ways, causing traditional n-gram–based matching metrics to over-penalize reasonable rewriting. In addition, these metrics cannot properly distinguish the importance of informative keywords from non-informative tokens like stop words. Furthermore, in real-world reverse engineering scenarios, it is often difficult to obtain high-quality and reliable reference texts.

\subsubsection{Decompilation Optimization (DO)} 
Compared to the first two tasks, this task emphasizes the structural quality and readability of the code itself. It aims to post-edit decompiler (e.g., IDA Pro) outputs through transformations including redundancy elimination, variable and type enhancement, and statement restructuring, making the result closer to source code conventions and style. Recent work, including LLM4Decompile~\cite{tan2024llm4decompile}, SK${^2}$Decompile~\cite{tan2025sk2decompile}, and ReCopilot~\cite{chen2025recopilot}, demonstrates the potential of LLMs for such Code2Code transformations, improving the readability of decompiled artifacts.
Although the Re-executability Rate is commonly used for evaluation, it only reflects functional correctness and cannot capture non-functional properties such as readability. Moreover, real-world binaries often lack unit tests and have hard-to-reproduce environments, which makes dynamic test-based metrics ineffective in in-the-wild settings, while match-based metrics still struggle with semantically equivalent but differently expressed outputs.

\vspace{-1ex}
\subsection{LLM-as-a-Judge in SE}\label{sec:llmjudgese}
The widespread adoption of LLMs in software engineering has led to the generation of a large volume of automated software artifacts, and the difficulty of evaluating such massive outputs has in turn given rise to the LLM-as-a-Judge paradigm~\cite{he2026llm, zhou2025se}. 
This paradigm extends the role of LLMs from mere artifact generation to artifact quality assessment, and it has been proven to achieve high agreement with human experts across a variety of scenarios, including requirements engineering~\cite{ahmed2025can,lubos2024leveraging}, code generation~\cite{yang2025code,zhuo2024ice,tong2024codejudge,weyssowcodeultrafeedback}, and software maintenance~\cite{kumar2024llms,jiang2025codejudgebench,zeng2025evaluating}, thus becoming a scalable and cost-effective alternative to manual review, and, by virtue of its ability to operate without reference texts and to perform deep semantic understanding, it offers inherent advantages over traditional automated metrics.

As discussed earlier, this paradigm is naturally suited for HOBRE evaluation. However, there is still a lack of systematic empirical studies, benchmark datasets, and task-specific judging models for the HOBRE setting, and this critical gap motivates our research.


%

\vspace{-1ex}
\section{\NoCaseChange{\bench} Benchmark}\label{sec:bench}
To systematically evaluate the effectiveness of LLM-as-a-Judge on the HOBRE task and its alignment with human expert judgment, we construct \bench, the first HOBRE-oriented evaluation benchmark, meticulously annotated by reverse engineering experts. 

\vspace{-1ex}
\subsection{Datasets Collection and Preprocessing}\label{sec:collect}
To ensure both realism and diversity, we first select 51 representative projects from the GNU~\cite{gnuftp} repository, which are widely adopted in binary reverse engineering research \cite{shang2024far,zhu2025misum,shang2025foc}. We then compile these source projects with GCC 8.2.0 compiler while preserving DWARF debugging information, producing 24 distinct binary variants that span 6 target architectures (\texttt{ARM\_32}, \texttt{ARM\_64}, \texttt{X86}, \texttt{X64}, \texttt{MIPS\_32}, and \texttt{MIPS\_64}) and 4 optimization levels (\texttt{-O0}, \texttt{-O1}, \texttt{-O2}, and \texttt{-O3}). For each compiled binary, we retain both the original version with debugging information and its corresponding stripped version, in which symbols and debug sections are removed. 

Subsequently, we employ IDA Pro~\cite{IDA} to decompile both the original and stripped binaries, and extract function boundaries, names, and corresponding pseudocode at the function level. While the stripped binaries replace function names with placeholder addresses, the function boundaries remain consistent, and we leverage this to align and merge content extracted from the two versions. 

To further obtain information from the source-code side, we use the srcML~\cite{collard2013srcml} tool to parse the source code into XML format and extract developer-written comments located above function definitions and declarations. Based on the established function boundaries and name mappings, we then align the source function body, source function name, and developer comments with the stripped pseudocode. Through this process, each function sample is organized into a unified structured representation, including the stripped pseudocode, original (symbol-preserving) pseudocode, source function body, source function name, and available developer comments, etc. In total, we obtain 346,596 function-level samples.

\vspace{-1.5ex}
\subsection{Response Generation}\label{sec:response}
To evaluate the discriminative capability of LLM-as-a-Judge across diverse HOBRE outputs with varying quality and styles, we select 8 representative models for each task to generate corresponding responses for every sample. These models fall into 
three categories:

\setlength{\leftmargini}{9pt}
\begin{itemize}
\item \textbf{DL-based Binary-specific Models:} For each task, we select two models, including SymLM~\cite{jin2022symlm} and HexT5~\cite{xiong2023hext5} for function name recovery, BinT5~\cite{al2023extending} and CP-BCS~\cite{ye2023cp} for binary code summarization, and LLM4Decompile~\cite{tan2024llm4decompile} and SK${^2}$Decompile~\cite{tan2025sk2decompile} for decompilation optimization. These models are specifically designed or trained for their corresponding tasks.

\item \textbf{General-Purpose LLMs:} This category covers the typical closed-source model GPT-4o~\cite{achiam2023gpt}, the ultra-large-scale open-source model DeepSeek-V3.2 (685B)~\cite{liu2025deepseek}, as well as code-oriented lightweight models DeepSeek-Coder-V2-Lite-Instruct (16B)~\cite{zhu2024deepseek}, Qwen3-Coder-30B-A3B-Instruct~\cite{yang2025qwen3}, and CodeLlama-34B-Instruct~\cite{roziere2023code}.

\item \textbf{Binary-Domain LLMs:} Specifically ReCopilot~\cite{chen2025recopilot}, which is built upon general-purpose LLMs and undergo continued pre-training and instruction-tuning on binary-specific corpora to better adapt to binary reverse engineering tasks.
\end{itemize}

For each function sample, we input the stripped pseudocode into the above models and generate task-specific outputs, including function names, natural language summaries, or optimized pseudocode.  Regarding prompt templates, models in categories (1) and (3) follow their default settings, while models in category (2) adopt the settings from \cite{shang2024far}. It is worth emphasizing that this stage aims not to pursue optimal generation quality, but rather to cover diverse model sources, output qualities, and expressive styles. Ultimately, this process yields a total of 2,772,768 ($346,596 \times 8$) candidate responses for each task.

\vspace{-1ex}
\subsection{Manual Annotation}\label{sec:annotation}

\subsubsection{\textbf{Sampling Strategy.}} Given the prohibitively high cost of manual expert annotation, we design a rigorous sampling strategy to ensure both statistical significance and diversity of \bench. First, targeting a 95\% confidence level with a 5\% confidence interval, we randomly sample 385 samples from the pool of 2,772,768 candidate responses.
We then introduce an additional coverage constraint requiring the samples to span diverse combinations of "\emph{generative model—target architecture—optimization level}" (totaling $8\times6\times4=192$ triplets).
Since the initial random samples already cover 166 such combinations, we perform targeted supplementary sampling for an additional 26 samples, ultimately forming a set of 411 samples per task for annotation.

\subsubsection{\textbf{Multi-dimensional Evaluation Framework.}} We employ a five-point multi-dimensional Likert scale to assess the overall quality of candidate outputs, incorporating a two-tier evaluation framework composed of shared and task-specific dimensions. Specifically, \textbf{(i) shared dimensions} are applicable to all three HOBRE tasks, aiming to ensure \emph{Semantic Faithfulness} (i.e., correctly reflecting the functionality of the pseudocode) and \emph{Analyst-centered Utility} (i.e., aligning with source-code-level representations to reduce cognitive burden); and \textbf{(ii) task-specific dimensions}, i.e., \emph{Naming Distinctiveness \& Naturalness} for function names, \emph{Information Coverage \& Density} for summaries, and \emph{Idiomization \& Faithfulness} for optimized pseudocode, are designed to capture the unique quality characteristics of different output forms. Detailed definitions of each dimension and the corresponding scoring criteria are provided in the human annotation guideline shown in Table~\ref{tab:guideline}. Overall, this framework comprehensively describes the quality implications of the HOBRE task.

\subsubsection{\textbf{Human Annotation and Quality Assurance.}} The three authors of this paper, each with over three years of experience in binary reverse engineering, served as annotation experts.
Prior to the formal annotation, the three experts jointly annotate 10 samples for each task and discuss the scoring criteria based on the annotation guideline in Table~\ref{tab:guideline} to calibrate their judgments and ensure a consistent understanding. 
Subsequently, the remaining samples are annotated independently, with the entire process spanning approximately two weeks per expert.
Upon completion, we measure the inter-annotator agreement among the three experts using ordinal Krippendorff's Alpha.
The coefficients for the three tasks reach 0.7996, 0.7077, and 0.6619, respectively, indicating a satisfactory level of agreement and confirming the reliability of the initial annotations.

To further improve annotation quality and mitigate subjective bias, we convene review meetings for samples with substantial disagreements (i.e., where the range between the maximum and minimum scores among the three experts is $\ge 2$, e.g., 3, 4, 5) to re-evaluate and establish final scores. Across the three tasks, 73, 105, and 76 samples undergo this review process, respectively.
For samples with minor disagreements (i.e., a score range < 2), we directly use the mode of the three scores as the final score (e.g., 4, 4, 5 → 4). The resulting expert scores are treated as the "gold standard" in \bench and are used to measure the consistency of subsequent automated evaluation methods with human judgment.

\begin{figure}[t]
	\centering
        \scalebox{1}{
	\includegraphics[width=1\linewidth]{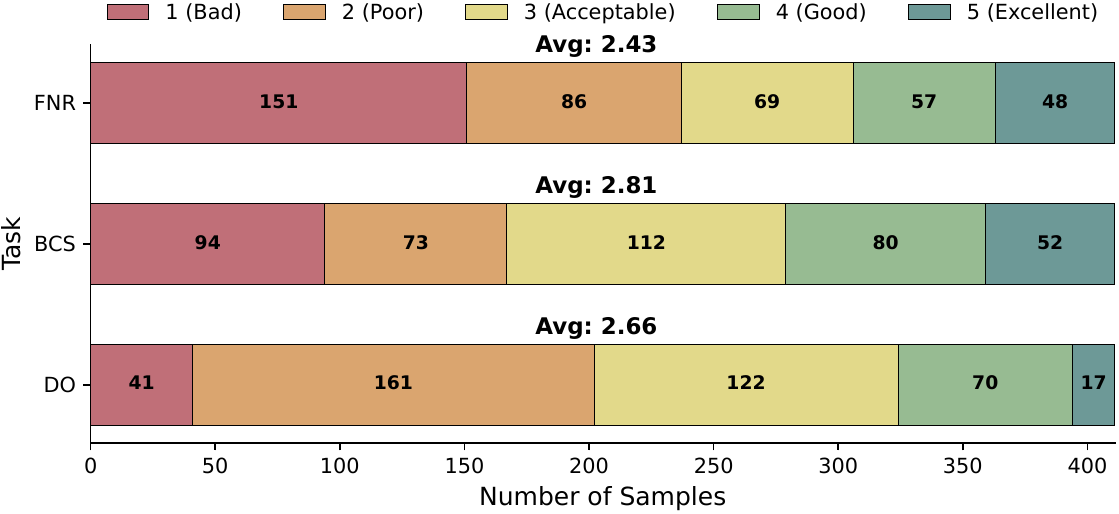}
        }
    \vspace{-4ex}
	\caption{Distribution of human expert-annotated scores across three HOBRE tasks in \bench.}
    \vspace{-2ex}
    \label{fig:distribution}
\end{figure}

\vspace{-1ex}
\subsection{Dataset Analysis}\label{sec:analysis}

Figure~\ref{fig:distribution} illustrates the distribution of expert annotation scores in \bench. The results indicate that \bench covers the full score spectrum from low quality (score 1) to high quality (score 5), demonstrating a well-defined gradient and discriminability. Among the three tasks, function name recovery receives the lowest average score (2.42), with the largest proportion of low-scoring samples; binary code summarization exhibits the most balanced distribution and the highest average score (2.81); while decompilation optimization obtains an average score of 2.66 and exhibits a clear scarcity of high-scoring samples. Overall, \bench provides reasonable quality coverage and sample diversity, serving as a reliable benchmark for subsequent evaluation.

\vspace{-1ex}
\section{Empirical Study}\label{sec:empirical}

Our empirical study investigates the following research questions to explore the effectiveness of LLM-as-a-Judge on the HOBRE task:

\setlength{\leftmargini}{9pt}
\begin{itemize}
\item \textbf{RQ1: Can LLM-as-a-Judge better approximate human expert judgment in HOBRE tasks?} We compare the performance of LLM-as-a-Judge against 12 traditional match-based and embedding-based automated metrics, verifying its higher correlation with human expert judgment.

\item \textbf{RQ2: How do different configurations of LLM-as-a-Judge affect evaluation correlation and cost?} We analyze three primary factors: (1) diverse backbone LLMs; (2) prompting strategies, i.e., Zero-Shot Learning (ZSL), Few-Shot Learning (FSL), and Chain-of-Thought (CoT); and (3) decoding strategies, specifically varying temperature. We explore the correlation with human judgment and API cost overhead under each configuration.

\item \textbf{RQ3: Does the optimal configuration of LLM-as-a-Judge vary across tasks or samples?} We investigate whether a "one-size-fits-all" configuration exists, i.e., whether the optimal configuration, which ensures high correlation while minimizing cost, varies across different tasks or sample characteristics.
\end{itemize}

\vspace{-1ex}
\subsection{Experimental Setup}
\subsubsection{\textbf{Environments}}
Our experimental environment operates on an Ubuntu 22.04 server, equipped with two 28-core Intel Xeon 6330 CPUs, 512GB RAM, 64TB storage, and 8*NVIDIA RTX A6000 GPUs, each with 48GB of VRAM. These GPUs run NVIDIA driver version 525.116.03 along with CUDA version 12.0. All LLMs used in this study are accessed via API calls through the OpenRouter~\cite{openrouter} platform, while all other experiments are deployed and trained locally on the above experimental environment.

\begin{table}[t]
  \centering
  \renewcommand{\arraystretch}{1.1}
  \caption{The Detailed Information of LLMs Employed in the LLM-as-a-Judge Evaluation.}
  \vspace{-2.0ex}
  \hspace*{-0.3cm} 
  \setlength{\tabcolsep}{0.63mm}{
  \scalebox{0.65}{
  \begin{threeparttable}
    \begin{tabular}{lclccccc}
    \toprule
        \multicolumn{1}{l}{\textbf{Model}} & \textbf{Size} & \multicolumn{1}{c}{\textbf{Model ID}} & \textbf{\makecell{Input\tnote{1} \\ Tokens}} & \textbf{\makecell{Output\tnote{1} \\ Tokens}} & \textbf{Publisher} & \textbf{Usage\tnote{2}} \\
    \midrule
        \multicolumn{5}{l}{\textbf{Proprietary LLMs}} \\
        GPT-4o~\cite{achiam2023gpt} & - & GPT-4o-2024-11-20 & \$2.50/M & \$10/M & OpenAI & $\bullet$ \\
        Claude-3.5~\cite{claude35_sonnet} & - & Claude-3.5-Sonnet & \$6/M & \$30/M & Anthropic & $\circ$\\
        Gemini-2.5~\cite{comanici2025gemini} & - & Gemini-2.5-Flash & \$0.30/M & \$2.50/M & Google & $\circ$ \\
        \cmidrule{1-7}
        \multicolumn{5}{l}{\textbf{Open-Source General LLMs}} \\
        DeepSeek-V3.2~\cite{liu2025deepseek} & 685B & DeepSeek-V3.2 & \$0.26/M & \$0.38/M & DeepSeek & $\bullet$ \\
        Mistral-Large-3~\cite{mistral2025large3} & 675B & Mistral-Large-3-675B-Instruct-2512  & \$0.50/M  & \$1.50/M & Mistral AI & $\circ$ \\
        Llama-3.3~\cite{grattafiori2024llama} & 70B & Llama-3.3-70B-Instruct & \$0.10/M  & \$0.32/M & Meta AI & $\circ$ \\
        Phi-4~\cite{abdin2024phi} & 14B & Phi-4 & \$0.065/M  & \$0.14/M & Microsoft & $\circ$ \\
        \cmidrule{1-7}
        \multicolumn{5}{l}{\textbf{Open-Source Code LLMs}} \\
        Qwen3-Coder~\cite{yang2025qwen3} & 30B & Qwen3-Coder-30B-A3B-Instruct & \$0.07/M & \$0.27/M  & Qwen & $\bullet$ \\ 
        Codestral~\cite{codestral2508} & - & Codestral-2508 & \$0.30/M & \$0.90/M & Mistral AI & $\circ$ \\ 
    \bottomrule
    \end{tabular}
    \begin{tablenotes}
        \normalsize 
        \item[1] Token prices (per million tokens) are sourced from OpenRouter~\cite{openrouter} as of March 22, 2026. 
        \item[2] "$\bullet$" indicates that the model was used in Response Generation (\S\ref{sec:response}), "$\circ$" indicates it was not. 
    \end{tablenotes}
    \end{threeparttable}}}
    \label{tab:model} 
    \vspace{-3ex}
\end{table}

\subsubsection{\textbf{Baseline Metrics.}} We select the following 2 categories, totaling 11 traditional automated evaluation metrics, as baselines:

\textbf{Match-based Metrics.} 
These metrics assess quality by measuring the surface-level overlap between generated text and references. BLEU~\cite{papineni2002bleu} and Rouge-L~\cite{lin-2004-rouge} capture lexical similarity based on $n$-gram overlaps and the longest common subsequence, respectively. METEOR~\cite{banerjee2005meteor} improves upon basic $n$-gram matching by considering stemmed forms, synonym matches, and word order, while ChrF++~\cite{popovic2017chrf} further assesses similarity using character-level $n$-gram precision and recall. CrystalBLEU~\cite{eghbali2022crystalbleu} filters out the most frequently occurring $n$-grams before computing BLEU to better capture substantive differences between generated content and reference. CodeBLEU~\cite{ren2020codebleu} enhances traditional $n$-gram matching by incorporating structural features like abstract syntax trees and data-flow dependencies. RUBY~\cite{tran2019does} evaluates similarity by jointly considering lexical, syntactic, and semantic representations of code.
F1-score is also adopted in some function name recovery works~\cite{gao2021lightweight}, which compute token-level F1-score by ignoring non-alphabetical characters and being case-, order-, and duplication-insensitive at the token level, and we follow their practice.

\textbf{Embedding-based Metrics.} These metrics map the generated text and the reference text into high-dimensional semantic spaces, measuring their semantic proximity via vector similarity. SentenceBERT~\cite{reimers2019sentence} employs the cosine similarity of sentence-level embeddings to assess overall semantic similarity. MoverScore~\cite{zhao2019moverscore} represents text as sequences of token embeddings and measures semantic divergence using Earth Mover's Distance. BERTScore~\cite{zhang2019bertscore} computes token-level similarity using contextual representations from a pre-trained BERT model, while CodeBERTScore~\cite{zhou2023codebertscore} adapts BERTScore specifically for code-related tasks to enhance its representation capabilities for programming languages.

\begin{table*}[t]
    \caption{Correlation of traditional metrics and LLM-as-a-Judge with human expert judgment across HOBRE tasks in \bench. The highest correlation is highlighted in bold and underlined, and the highest in traditional metrics is underlined.}
    \vspace{-2.4ex}
    \centering
    \renewcommand{\arraystretch}{0.95}
    \setlength{\tabcolsep}{1mm}
    \scalebox{0.88}{
    \begin{threeparttable}
    \begin{tabular}{lC{1.41cm}C{1.41cm}C{1.41cm}C{1.41cm}C{1.41cm}C{1.41cm}C{1.41cm}C{1.41cm}C{1.41cm}c}
        \toprule
            \multirow{2}{*}{\textbf{Metrics}} & \multicolumn{3}{c}{\textbf{Function Name Recovery (FNR)}} & \multicolumn{3}{c}{\textbf{Binary Code Summarization (BCS)}} & \multicolumn{3}{c}{\textbf{Decompilation Optimization (DO)}} & \multirow{2}{*}{\textbf{Avg.}}   \\
        \cmidrule(lr){2-4}  \cmidrule(lr){5-7}  \cmidrule(lr){8-10}  
            & \textbf{$\tau$ (\%)} & \textbf{$r_s$ (\%)}  & \textbf{$r_p$ (\%)} & \textbf{$\tau$ (\%)} & \textbf{$r_s$ (\%)}  & \textbf{$r_p$ (\%)} & \textbf{$\tau$ (\%)} & \textbf{$r_s$ (\%)}  & \textbf{$r_p$ (\%)} & \\
        \midrule
            \rowcolor{gray!18} \multicolumn{11}{c}{\textbf{Existing Match-based / Embedding-based Metrics}} \\
        \midrule
            F1-score     & 48.51 & 54.65   & 60.15   & --   & --   & --   & --   & --   & --   & --\\
            BLEU         & 31.58 & 40.18 & 50.69 & 26.89 & 35.45 & 27.91 & 25.80 & 33.01 & 33.50 & 33.89\\
            METEOR       & \underline{49.97} & 56.79 & 57.36 & 27.89 & 36.45 & 34.25 & 24.67 & 31.95 & 32.29 & 39.05 \\
            ROUGE-L      & 48.29 & 54.51 & 59.51 & 31.12 & 40.03 & 38.73 & 28.63 & 37.05 & 39.48 & 41.93\\
            ChrF++       & 47.55 & \underline{60.37} & 65.24 & 35.32 & 46.88 & 46.64 & \underline{29.24} & \underline{37.63} & \underline{40.60} & 45.40\\
            CrystalBLEU  & 32.60 & 42.19 & 53.12 & 19.80 & 26.77 & 25.66 & 24.49 & 31.30 & 29.55 & 33.06\\
            CodeBLEU     & 13.08 & 14.49 & \hspace{1ex}8.65 & 12.49 & 16.27 & 00.74 & 16.62 & 22.02 & 23.29 & 14.18 \\
            RUBY     & 24.28 & 26.83 & 30.74 & 27.88 & 36.02 & 31.73 & 16.78 & 21.78 & 25.87 & 26.88 \\
        \midrule
            SentenceBERT & 43.86 & 57.02 & \underline{65.54} & \underline{41.67} & \underline{54.17} & \underline{56.66} & 28.10 & 36.51 & 38.19 & \underline{46.86} \\
            MoverScore   & 31.43 & 40.92 & 53.49 & 28.41 & 38.14 & 38.01 & 11.92 & 15.98 & 18.68 & 30.78 \\
            BERTScore    & 36.03 & 47.16 & 51.45 & 28.62 & 38.18 & 36.95 & 22.21 & 29.01 & 33.21 & 35.87 \\
            CodeBERTScore& 38.25 & 49.15 & 51.32 & 25.26 & 33.67 & 34.82 & 28.13 & 36.40 & 40.14 & 37.46 \\
        \cmidrule(lr){1-11}
            \textbf{Average} & 37.12 & 45.36 & 50.61 & 27.76 & 36.55 & 33.83 & 23.55 & 30.53 & 32.82 & 35.04 \\
        \midrule
            \rowcolor{gray!18} \multicolumn{11}{c}{\textbf{LLM-as-a-Judge Paradigm}} \\
        \midrule
            Codestral-2508         & 55.28 & 65.02 & 60.83 & 48.44 & 58.40 & 57.63 & 40.38 & 45.45 & 48.28 & 53.30 \\
            Phi-4                  & 48.64 & 56.50 & 55.23 & 56.89 & 65.40 & 65.77 & 41.68 & 47.22 & 50.20 & 54.17 \\
            Llama-3.3-70B & 60.97 & 70.04 & 64.42 & 61.27 & 68.88 & 69.78 & 48.14 & 52.82 & 52.47 & 60.97 \\
            Qwen3-Coder-30B        & 63.64 & 72.85 & 70.01 & 65.68 & 75.09 & 74.13 & 42.82 & 47.83 & 48.78 & 62.32  \\
            Mistral-Large-3   & 59.28 & 68.71 & 66.95 & 63.96 & 72.71 & 73.02 & 53.95 & 59.32 & 59.89 & 64.20 \\
            GPT-4o                 & 64.44 & 73.86 & 69.57 & 65.94 & 75.26 & 75.43 & 51.17 & 57.55 & 57.38 & 65.62 \\
            Claude-3.5-Sonnet      & 67.04 & 75.66 & 73.33 & 67.46 & 76.44 & 76.98 & 54.70 & 61.10 & 61.98 & 68.30 \\
            Gemini-2.5-Flash       & \underline{\textbf{69.02}} & \underline{\textbf{78.92}} & \underline{\textbf{75.93}} & \underline{\textbf{68.09}} & \underline{\textbf{77.45}} & \underline{\textbf{77.90}} & 52.76 & 60.28 & 60.94 & 69.03 \\
            DeepSeek-V3.2          & 66.17 & 75.03 & 74.28 & 67.36 & 76.58 & 76.68 & \underline{\textbf{62.11}} & \underline{\textbf{69.75}} & \underline{\textbf{69.88}} & \underline{\textbf{70.87}} \\
        \cmidrule(lr){1-11}
            \textbf{Average} & 61.61 & 70.73 & 67.84 & 62.79 & 71.80 & 71.92 & 49.74 & 55.70 & 56.65 & 63.20 \\
        \bottomrule
    \end{tabular} 
    \end{threeparttable} } 
    \label{tab:overall}
    \vspace{-1.6ex}
\end{table*}

\subsubsection{\textbf{Meta-Evaluation.}}
To measure the consistency between various automated evaluation metrics (including both traditional baselines and the LLM-as-a-Judge paradigm) and human expert judgment, we employ three statistical correlation coefficients widely used in prior research~\cite{zhuo2024ice, wang2025can, ouedraogo2025human,dong2025codescore}, i.e., \emph{Kendall’s $\tau$ coefficient}, \emph{Spearman’s} $r_s$, and \emph{Pearson’s} $r_p$, to quantify the association between automated scores and the expert-annotated "gold standard" in \bench. Specifically, Kendall’s $\tau$ coefficient~\cite{kendall1938new} measures ordinal associations between sample pairs, Spearman’s $r_s$~\cite{spearman1961proof} assesses the monotonic correlation between their ranking orders, and Pearson’s $r_p$~\cite{pearson1896vii} measures the linear correlation between them. 
Considering that expert ratings are ordinal discrete values (on a 1–5 scale), we prioritize Kendall’s $\tau$ and Spearman’s $r_s$ as the primary reported metrics, while treating Pearson’s $r_p$ as a supplementary reference. Taken together, these three coefficients provide a meta-evaluation of automated evaluation metrics.

\begin{table*}[t]
    \caption{Correlation of various LLM-as-a-Judge configurations (including different LLMs, prompting strategies, and decoding temperatures) with human expert judgment in \bench. Each cell reports the "Kendall’s $\tau$ coefficient (\%) / API cost (\$)". For each LLMs, the highest correlation within its $3 \times 3$ sub-block is highlighted in bold and underlined.}
    \vspace{-2ex}
    \centering
    \hspace*{-4.6cm} 
    \renewcommand{\arraystretch}{0.95}
    \setlength{\tabcolsep}{1.65mm}
    \scalebox{0.74}{
    \begin{threeparttable}
    \begin{NiceTabular}{lcccc|ccc|ccc|c}
        \toprule
            \omit \multirow{2}{*}{\textbf{Models}} & \multirow{2}{*}{\textbf{Prompt}}  & \multicolumn{3}{c}{\textbf{Function Name Recovery (FNR)}} & \multicolumn{3}{c}{\textbf{Binary Code Summarization (BCS)}} & \multicolumn{3}{c}{\textbf{Decompilation Optimization (DO)}} & \multirow{2}{*}{\textbf{Average}}  \\
        \cmidrule(lr){3-5}  \cmidrule(lr){6-8}  \cmidrule(lr){9-11}  
            \omit &  & T=0.1 & T=0.5 & T=1.0 & T=0.1 & T=0.5 & T=1.0 & T=0.1 & T=0.5 & T=1.0 & \\
        \midrule
            \multicolumn{1}{l}{\multirow{3}{*}{Codestral-2508}} 
                & ZSL & 45.44 / 0.100 & 46.61 / 0.100 & 44.71 / 0.100 & \underline{\textbf{53.15}} / 0.106 & 51.67 / 0.106 & 47.56 / 0.106 & 39.81 / 0.152 & \underline{\textbf{41.63}} / 0.152 & 29.29 / 0.152 & 44.43 / 0.119 \\
                & FSL & 31.70 / 0.322 & 30.28 / 0.322 & 29.22 / 0.322 & 49.35 / 0.238 & 44.87 / 0.238 & 43.95 / 0.238 & 35.08 / 0.494 & 32.32 / 0.494 & 34.97 / 0.494 & 36.86 / 0.351 \\
                & CoT & 55.28 / 0.235 & 51.12 / 0.237 & \underline{\textbf{56.98}} / 0.237 & 48.44 / 0.218 & 48.33 / 0.220 & 46.04 / 0.220 & 40.38 / 0.294 & 38.11 / 0.292 & 36.29 / 0.293 & \underline{\textbf{46.77}} / 0.250 \\
            \cmidrule(lr){2-12}
            \multicolumn{1}{l}{\multirow{3}{*}{Phi-4}} 
                & ZSL & 47.29 / 0.020 & 42.31 / 0.020 & 37.68 / 0.020 & 47.57 / 0.022 & 48.14 / 0.022 & 48.22 / 0.022 & 30.47 / 0.031 & \underline{\textbf{42.84}} / 0.030 & 25.13 / 0.031 & 41.07 / 0.024 \\
                & FSL & 36.69 / 0.066 & 33.25 / 0.066 & 48.64 / 0.065 & 45.02 / 0.048 & 42.85 / 0.048 & \underline{\textbf{56.89}} / 0.048 & 36.67 / 0.101 & 28.54 / 0.101 & 41.68 / 0.100 & 41.14 / 0.071 \\
                & CoT & 42.80 / 0.044 & 40.38 / 0.045 & \underline{\textbf{49.66}} / 0.048 & 51.61 / 0.042 & 50.33 / 0.043 & 54.74 / 0.045 & 35.56 / 0.056 & 35.72 / 0.057 & 32.18 / 0.059 & \underline{\textbf{43.66}} / 0.049 \\
            \cmidrule(lr){2-12}
            \multicolumn{1}{l}{\multirow{3}{*}{Llama-3.3-70B}} 
                & ZSL & 60.97 / 0.032 & 61.20 / 0.032 & \underline{\textbf{61.82}} / 0.032 & 61.27 / 0.034 & 60.30 / 0.034 & 62.36 / 0.034 & \underline{\textbf{48.14}} / 0.048 & 45.96 / 0.048 & 43.54 / 0.047 & \underline{\textbf{56.17}} / 0.038 \\
                & FSL & 59.53 / 0.101 & 58.33 / 0.101 & 59.61 / 0.102 & \underline{\textbf{62.79}} / 0.074 & 61.50 / 0.074 & 61.53 / 0.075 & 47.80 / 0.155 & 46.17 / 0.155 & 41.63 / 0.155 & 55.43 / 0.110 \\
                & CoT & 53.50 / 0.094 & 49.74 / 0.093 & 52.73 / 0.094 & 58.08 / 0.089 & 59.79 / 0.090 & 57.01 / 0.090 & 46.05 / 0.109 & 45.89 / 0.111 & 46.45 / 0.108 & 52.14 / 0.097 \\
            \cmidrule(lr){2-12}
            \multicolumn{1}{l}{\multirow{3}{*}{Qwen3-Coder-30B}} 
                & ZSL & 64.00 / 0.023 & 63.20 / 0.023 & 62.40 / 0.023 & 63.16 / 0.025 & 63.07 / 0.025 & 63.47 / 0.025 & 38.06 / 0.035 & 38.67 / 0.035 & 39.35 / 0.035 & 55.04 / 0.028 \\
                & FSL & 63.64 / 0.075 & \underline{\textbf{64.28}} / 0.075 & 62.73 / 0.075 & 65.68 / 0.056 & 64.33 / 0.056 & \underline{\textbf{66.96}} / 0.056 & 42.82 / 0.115 & 41.84 / 0.115 & 40.28 / 0.115 & \underline{\textbf{56.95}} / 0.082 \\
                & CoT & 56.25 / 0.079 & 53.37 / 0.078 & 49.71 / 0.083 & 55.43 / 0.075 & 54.00 / 0.076 & 55.29 / 0.081 & \underline{\textbf{43.68}} / 0.100 & 41.38 / 0.101 & 38.83 / 0.108 & 49.77 / 0.087 \\
            \cmidrule(lr){2-12}
            \multicolumn{1}{l}{\multirow{3}{*}{Mistral-Large-3}} 
                & ZSL & 63.40 / 0.167 & \underline{\textbf{64.09}} / 0.167 & 63.11 / 0.167 & 58.25 / 0.177 & 60.31 / 0.177 & 58.12 / 0.177 & 50.08 / 0.254 & 51.95 / 0.254 & 51.23 / 0.254 & 57.84 / 0.199 \\
                & FSL & 59.98 / 0.537 & 59.28 / 0.537 & 59.72 / 0.537 & \underline{\textbf{64.23}} / 0.396 & 63.96 / 0.396 & 63.98 / 0.396 & 52.84 / 0.824 & \underline{\textbf{53.95}} / 0.824 & 52.80 / 0.824 & \underline{\textbf{58.97}} / 0.586 \\
                & CoT & 53.98 / 0.570 & 53.77 / 0.570 & 54.93 / 0.570 & 61.02 / 0.488 & 58.00 / 0.494 & 59.40 / 0.490 & 45.88 / 0.696 & 49.53 / 0.695 & 45.92 / 0.683 & 53.60 / 0.584 \\
            \cmidrule(lr){2-12}
            \multicolumn{1}{l}{\multirow{3}{*}{GPT-4o}} 
                & ZSL & 65.53 / 0.796 & 65.78 / 0.796 & 65.08 / 0.796 & 60.48 / 0.845 & 60.70 / 0.845 & 59.63 / 0.845 & 47.51 / 1.181 & 46.32 / 1.181 & 46.04 / 1.181 & 57.45 / 0.941 \\
                & FSL & 64.44 / 2.533 & 64.29 / 2.533 & 63.33 / 2.533 & \underline{\textbf{65.94}} / 1.856 & 64.82 / 1.856 & 63.48 / 1.856 & \underline{\textbf{51.17}} / 3.875 & 50.19 / 3.875 & 47.96 / 3.875 & \underline{\textbf{59.51}} / 2.754 \\
                & CoT & \underline{\textbf{65.95}} / 3.075 & 64.64 / 3.112 & 61.76 / 3.267 & 58.11 / 2.909 & 61.90 / 3.011 & 57.14 / 3.256 & 50.88 / 4.288 & 51.58 / 4.148 & 50.64 / 4.281 & 58.07 / 3.483 \\
            \cmidrule(lr){2-12}
            \multicolumn{1}{l}{\multirow{3}{*}{Claude-3.5-Sonnet}} 
                & ZSL & \underline{\textbf{70.31}} / 2.333 & 69.11 / 2.333 & 66.93 / 2.332 & 60.23 / 2.465 & 59.44 / 2.465 & 60.84 / 2.464 & \underline{\textbf{57.83}} / 3.526 & 56.59 / 3.525 & 53.55 / 3.525 & \underline{\textbf{61.65}} / 2.774 \\
                & FSL & 67.04 / 7.606 & 57.39 / 7.616 & 55.21 / 7.636 & \underline{\textbf{67.46}} / 5.516 & 64.41 / 5.546 & 63.37 / 5.596 & 54.70 / 11.69 & 52.13 / 11.70 & 48.74 / 11.71 & 58.94 / 8.290 \\
                & CoT & 61.17 / 6.095 & 56.77 / 6.091 & 55.73 / 6.071 & 56.08 / 5.730 & 57.08 / 5.722 & 56.34 / 5.759 & 50.75 / 7.302 & 45.95 / 7.220 & 42.07 / 7.057 & 53.55 / 6.339 \\
            \cmidrule(lr){2-12}
            \multicolumn{1}{l}{\multirow{3}{*}{Gemini-2.5-Flash}} 
                & ZSL & 66.56 / 0.110 & 66.61 / 0.110 & 68.31 / 0.110 & 66.36 / 0.116 & 66.83 / 0.116 & 66.40 / 0.116 & 49.11 / 0.167 & 50.15 / 0.167 & 51.94 / 0.167 & 61.36 / 0.131 \\
                & FSL & \underline{\textbf{69.02}} / 0.355 & 68.20 / 0.355 & 65.81 / 0.355 & 68.09 / 0.261 & \underline{\textbf{68.13}} / 0.261 & 64.90 / 0.261 & \underline{\textbf{52.76}} / 0.550 & 51.88 / 0.550 & 48.50 / 0.550 & \underline{\textbf{61.92}} / 0.389 \\
                & CoT & 57.23 / 1.181 & 56.12 / 1.252 & 56.98 / 1.313 & 60.77 / 1.436 & 61.19 / 1.500 & 61.60 / 1.542 & 39.96 / 2.314 & 38.76 / 2.369 & 33.77 / 2.663 & 51.82 / 1.730 \\
            \cmidrule(lr){2-12}
            \multicolumn{1}{l}{\multirow{3}{*}{DeepSeek-V3.2}} 
                & ZSL & 63.96 / 0.083 & 63.72 / 0.083 & 61.61 / 0.083 & 63.46 / 0.088 & 63.52 / 0.088 & 61.38 / 0.088 & 53.51 / 0.125 & 53.89 / 0.125 & 51.72 / 0.124 & 59.64 / 0.099 \\
                & FSL & \underline{\textbf{66.17}} / 0.267 & 62.76 / 0.267 & 62.82 / 0.267 & 67.36 / 0.197 & \underline{\textbf{68.18}} / 0.197 & 61.00 / 0.197 & \underline{\textbf{62.11}} / 0.411 & 57.25 / 0.411 & 57.17 / 0.411 & \underline{\textbf{62.76}} / 0.291 \\
                & CoT & 57.90 / 0.169 & 58.81 / 0.164 & 55.76 / 0.168 & 59.82 / 0.164 & 63.38 / 0.165 & 62.39 / 0.170 & 53.49 / 0.219 & 51.14 / 0.221 & 50.97 / 0.228 & 57.07 / 0.185 \\
        \cmidrule(lr){1-12}
            \omit  \textbf{Average} & & \underline{\textbf{58.14}} / 1.003 & 56.50 / 1.007 & 56.78 / 1.015 & \underline{\textbf{59.23}} / 0.877 & 58.93 / 0.884 & 58.67 / 0.898 & \underline{\textbf{46.56}} / 1.449 & 45.94 / 1.443 & 43.80 / 1.453 & 53.84 / 1.114 \\
        \bottomrule
    \end{NiceTabular} 
    \end{threeparttable} } 
    \label{tab:config}
    \vspace{-1.6ex}
\end{table*}

\vspace{-1.2ex}
\subsection{RQ1: Meta-Evaluation of LLM-as-a-Judge} \label{sec:compare}

\textbf{Evaluation Setup.} To comprehensively evaluate the effectiveness of the LLM-as-a-Judge paradigm, we select 9 representative backbone LLMs, with detailed information provided in Table~\ref{tab:model}. These models span three categories—Proprietary, Open-Source General, and Open-Source Code—and encompass a range of parameter scales. Notably, within each category, we included one model that is previously used for response generation in \S\ref{sec:response} (i.e., GPT-4o, DeepSeek-V3.2, and Qwen3-Coder), aiming to investigate whether LLMs exhibit self-evaluation bias when judging HOBRE tasks.

Subsequently, we adopt the human expert annotations in \bench as the "gold standard" and compute Kendall’s $\tau$, Spearman’s $r_s$, and Pearson’s $r_p$ correlation coefficients for both 11 traditional automated metrics and the 9 backbone LLMs serving as judges. Due to space constraints, Table~\ref{tab:overall} reports only the configuration (i.e., prompting strategy and decoding temperature) that achieves the highest average correlation for each LLM across the three tasks. The impact of different parameter configurations on evaluation consistency will be discussed in detail in RQ2 (\S\ref{sec:config}).

\noindent\textbf{Superiority over Traditional Metrics.} As illustrated in Table~\ref{tab:overall}, the LLM-as-a-Judge paradigm achieves an average correlation of 63.20\% across all three HOBRE tasks, substantially outperforming the average of traditional metrics (35.04\%). Taking the FNR task as an example, the best-performing LLM, Gemini-2.5-Flash, reaches a Kendall’s $\tau$ of 69.02\%, nearly 1.4 times that of the best traditional metric, METEOR (49.97\%). This indicates that LLMs can transcend surface-level lexical matching to deeply comprehend functional semantics and expressive intent, yielding judgments that closely align with human experts.
While certain match-based metrics show marginal utility in specific scenarios (e.g., ChrF++ achieves 60.37\% $\tau$ in FNR, surpassing the lower-tier LLM Phi-4 at 56.50\%), they consistently fail to compete with high-parameter LLM evaluators.   
Among embedding-based metrics, SentenceBERT slightly outperforms a few lightweight LLMs on FNR in terms of Pearson’s $r_p$, yet remains below the median level of LLMs and exhibits even weaker performance on the BCS and DO tasks.

\noindent\textbf{Task-Specific Performance Variation.} Considering task types, LLMs perform best on the BCS task (average Kendall’s $\tau$ of 62.68\%), where the performance gap over traditional metrics is most pronounced (a 35.03\% difference in $\tau$). This can be attributed to the nature of code summarization as a typical open-ended generation task, which allows for numerous semantically equivalent expressions. In such cases, surface-level lexical overlap often decouples from actual semantic similarity, rendering $n$-gram matching or static embedding-based metrics ineffective.
In contrast, the performance gap narrows slightly in the DO task (average $\tau$ of 49.74\% for LLMs, and 23.55\% for traditional metrics). This suggests that the structured nature of Code2Code transformations allows traditional metrics to capture basic syntactic correctness. However, the critical advantage of LLMs on DO task lies in their ability to detect "superficially plausible but logically flawed" refactorings, which involves a deep semantic equivalence judgment that remains beyond the capability boundary of traditional metrics.

\noindent\textbf{Investigation of Self-Evaluation Bias.} We further analyze whether GPT-4o, DeepSeek-V3.2, and Qwen3-Coder, which serve as generators in our response generation stage, exhibit any bias when acting as evaluators. To investigate this, we partition the samples into $S_{\text{self}}$ (outputs generated by the LLM itself) and $S_{\text{others}}$ (outputs generated by other LLMs) subsets, and calculate the correlation of the LLM’s judgment with human experts for each subset. The results show that $\tau_{\text{self}} \approx \tau_{\text{others}}$, with the difference not reaching statistical significance, indicating that in the HOBRE evaluation context, LLMs maintain an objective evaluation scale and do not exhibit any preference toward their own generation style.


\noindent\textbf{Investigation of Hard Samples and the "Fluency Trap".} We define a "hard sample" as one with substantial disagreement between human experts and LLMs, i.e., $\vert{}\text{HumanScore} - \text{avg. LLMScore}\vert{} \ge 2$. Across \bench, 149 out of 1,233 samples (12.08\%) fall into this category (70/18/61 in FNR/BCS/DO). Task-stratified manual analysis on 100 hard samples reveals three primary difficulty patterns:
(1) Lack of semantic anchors (34\%): pseudocode lacking residual semantic clues (e.g., hard-coded strings or external API calls) forces LLM judges into speculative scoring.
(2) Lack of contextual information (14\%): isolated single-function snippets miss caller/callee relations, key global variables, or struct fields required for reliable judgment.
(3) Rewarding fluent-but-unsupported outputs (12\%): LLMs tend to over-score syntactically fluent or conventional outputs even if semantically unsupported.
To further quantify this "Fluency Trap", we treat samples with $\text{HumanScore} \le 2$ as "semantically unsupported," and from this set manually identify 100 seemingly fluent samples as a sub-pool.
In this fluent sub-pool, the proportion of LLM judge scores >3 rises to 59.0\% (compared to 46.7\% in the overall unsupported pool), which mainly comes from the FNR task, which tends to give medium scores (3–4) to plausible-looking names; for BCS and DO, the bias is not obvious. 
Overall, while LLM judges exhibit a slight bias toward fluency, its impact remains within a controllable and acceptable margin.

\begin{tcolorbox}[colback=gray!5,
                  colframe=black,
                  arc=0.8mm, auto outer arc,
                  boxrule=1pt,
                  boxsep=-2pt
                 ]
\textbf{Answering RQ1:} LLM-as-a-Judge achieves an average correlation of 63.20\% with human judgement, substantially superior traditional metrics at 35.04\%,  accurately capturing deep semantics while maintaining an objective scale, providing a high-fidelity, scalable automated evaluation solution.
\end{tcolorbox}

\begin{table*}[t]
    \caption{Comparison of Correlation and Cost Across Different Judge Configuration Selection Strategies ($\textcolor{customgreen}{\downarrow}$ and $\textcolor{customred}{\times}$ represent the correlation gap and cost ratio relative to the Oracle Configuration, respectively) (Due to space constraints, we ignore $r_p$).}
    \vspace{-2.2ex}
    \centering
    \renewcommand{\arraystretch}{1}
    \setlength{\tabcolsep}{1.1mm}
    \scalebox{0.89}{
    \begin{threeparttable}
    \begin{tabular}{lccC{1.41cm}C{1.41cm}C{1.41cm}C{1.41cm}C{1.41cm}C{1.41cm}C{1.41cm}C{1.41cm}C{1.41cm}}
        \toprule
             \multicolumn{3}{c}{\textbf{Configuration}} & \multicolumn{3}{c}{\textbf{Function Name Recovery (FNR)}} & \multicolumn{3}{c}{\textbf{Binary Code Summarization (BCS)}} & \multicolumn{3}{c}{\textbf{Decompilation Optimization (DO)}}   \\
        \cmidrule(lr){1-3} \cmidrule(lr){4-6}  \cmidrule(lr){7-9}  \cmidrule(lr){10-12}  
            Models & Prompt & Temp. & \textbf{$\tau$ (\%)} & \textbf{$r_s$ (\%)}  & Cost (\$) & \textbf{$\tau$ (\%)} & \textbf{$r_s$ (\%)}  & Cost (\$)  & \textbf{$\tau$ (\%)} & \textbf{$r_s$ (\%)}  & Cost (\$)  \\
        \midrule
        \multicolumn{3}{l}{\textbf{Random Configuration}} & $54.08_{\textcolor{customgreen}{(\downarrow41.8\%)}}$ & $63.56_{\textcolor{customgreen}{(\downarrow33.3\%)}}$ & $1.080_{\textcolor{customred}{(11.24\times)}}$ & $55.02_{\textcolor{customgreen}{(\downarrow44.1\%)}}$ & $64.76_{\textcolor{customgreen}{(\downarrow34.6\%)}}$ & $1.035_{\textcolor{customred}{(15.30\times)}}$ & $39.67_{\textcolor{customgreen}{(\downarrow56.1\%)}}$ & $45.82_{\textcolor{customgreen}{(\downarrow50.4\%)}}$ & $1.389_{\textcolor{customred}{(6.90\times)}}$ \\
        \cmidrule(lr){1-12}                    
            \multicolumn{3}{l}{\textbf{Static Configuration}} \\
            Claude-3.5-Sonnet & ZSL & 0.1 & $\textbf{70.31}_{\textcolor{customgreen}{(\downarrow24.4\%)}}$ & $\textbf{79.26}_{\textcolor{customgreen}{(\downarrow16.9\%)}}$ & $2.333_{\textcolor{customred}{(24.27\times)}}$ & $60.23_{\textcolor{customgreen}{(\downarrow38.8\%)}}$ & $69.47_{\textcolor{customgreen}{(\downarrow29.9\%)}}$ & $2.465_{\textcolor{customred}{(36.46\times)}}$ & $57.83_{\textcolor{customgreen}{(\downarrow36.0\%)}}$ & $64.88_{\textcolor{customgreen}{(\downarrow29.8\%)}}$ & $3.526_{\textcolor{customred}{(17.51\times)}}$  \\
            Gemini-2.5-Flash  & FSL & 0.5 & $68.20_{\textcolor{customgreen}{(\downarrow26.6\%)}}$ & $78.12_{\textcolor{customgreen}{(\downarrow18.1\%)}}$ & $0.355_{\textcolor{customred}{(3.69\times)}}$ & $\textbf{68.13}_{\textcolor{customgreen}{(\downarrow30.7\%)}}$ & $\textbf{77.66}_{\textcolor{customgreen}{(\downarrow21.6\%)}}$ & $0.261_{\textcolor{customred}{(3.86\times)}}$ & $51.88_{\textcolor{customgreen}{(\downarrow42.6\%)}}$ & $59.41_{\textcolor{customgreen}{(\downarrow35.7\%)}}$ & $0.550_{\textcolor{customred}{(2.73\times)}}$ \\
            DeepSeek-V3.2  & FSL & 0.1 & $66.17_{\textcolor{customgreen}{(\downarrow28.8\%)}}$ & $75.03_{\textcolor{customgreen}{(\downarrow21.3\%)}}$ & $0.267_{\textcolor{customred}{(2.78\times)}}$ & $67.36_{\textcolor{customgreen}{(\downarrow31.5\%)}}$ & $76.58_{\textcolor{customgreen}{(\downarrow22.7\%)}}$ & $0.197_{\textcolor{customred}{(2.91\times)}}$ & $\textbf{62.11}_{\textcolor{customgreen}{(\downarrow31.3\%)}}$ & $\textbf{69.75}_{\textcolor{customgreen}{(\downarrow24.5\%)}}$ & $0.411_{\textcolor{customred}{(2.04\times)}}$ \\
        \cmidrule(lr){1-12}
            \multicolumn{3}{l}{\textbf{Oracle Configuration}} & 92.95 & 95.33 & 0.096 & 98.36 & 99.02 & 0.068 & 90.41 & 92.38 & 0.201 \\
        \bottomrule
    \end{tabular} 
    \end{threeparttable} } 
    \vspace{-2.2ex}
    \label{tab:optimal}
\end{table*}

\vspace{-1ex}
\subsection{RQ2: Impact of Judge Configuration} \label{sec:config}

\textbf{Evaluation Setup.} To investigate the impact of different configurations (including backbone LLMs, prompting strategies, and decoding temperatures) on the correlation and overhead of LLM-as-a-Judge, we employ the 9 backbone LLMs listed in Table 2. Regarding prompting strategies, Zero-Shot Learning (ZSL) provides the LLMs with task descriptions, evaluation dimensions, scoring criteria, as well as the stripped decompiled pseudocode and candidate outputs, while Few-Shot Learning (FSL) builds upon this by providing one example for each score level (1–5). Chain-of-Thought (CoT) requires the LLMs to provide a step-by-step reasoning process before outputting the final score. Decoding temperatures are set to 0.1, 0.5, and 1.0 following prior work~\cite{ouedraogo2025human}. Due to space constraints, we report only Kendall’s $\tau$ and API costs in Table~\ref{tab:config}, as Spearman’s $r_s$ and Pearson’s $r_p$ exhibit trends consistent with $\tau$.

\noindent\textbf{Impact of Prompting Strategy.} 
The impact of the prompt strategy on correlation varies significantly across models of different scales. 
For ultra-large-scale LLMs (e.g., GPT-4o, Gemini-2.5-Flash, DeepSeek-V3.2, and Mistral-Large-3), FSL generally achieves the highest correlation. For instance, Gemini-2.5-Flash reaches a Kendall’s $\tau$ of 69.02\% in the FNR task at T=0.1 under FSL, outperforming ZSL (66.56\%) and CoT (57.23\%). This superiority stems from the ability of large LLMs to rapidly extract evaluation patterns from few-shot examples. However, the increased input tokens brought by FSL make its cost the highest, even exceeding that of CoT. Conversely, for small- to medium-scale LLMs (e.g., Codestral-2508 and Phi-4), FSL leads to a notable decline in correlation. This may be attributed to capacity constraints, where LLMs become "anchored" to specific patterns within the provided examples, thereby hindering their generalization capabilities.
In contrast, CoT provides a substantial boost for these smaller LLMs, achieving peak average $\tau$ values for Codestral-2508 (46.77\%) and Phi-4 (43.66\%). This suggests that explicit reasoning helps compensate for the weaker native reasoning capabilities of smaller LLMs. Interestingly, CoT often leads to a reduction in correlation for ultra-large LLMs in most cases, potentially because excessive reasoning induces hallucinations or over-interpretation, thereby interfering with judgment.

\noindent\textbf{Impact of Decoding Temperature.} 
Lower temperatures (0.1 or 0.5) generally yield higher correlation, particularly for ultra-large LLMs, as they help maintain a consistent evaluation scale in complex semantic judgments. For smaller LLMs, however, higher temperatures sometimes achieve optimal correlation under certain configurations. For example, Codestral-2508 and Phi-4 attain their highest correlation on the FNR task using CoT at T=1.0. This suggests that moderate randomness may help weaker LLMs escape local scoring traps and activate more diverse reasoning paths. Furthermore, increased temperature only results in slight cost growth under CoT due to higher variability in output length.

\noindent\textbf{Pareto Differences between LLMs.} 
Overall, correlation is positively associated with the LLM scale, accompanied by a marked increase in cost. However, within the ultra-large LLMs, substantial Pareto differences exist regarding evaluation efficiency and economy.
For example, on the FNR task using ZSL and T=0.1 configurations, Gemini-2.5-Flash and DeepSeek-V3.2 demonstrate extremely high cost-effectiveness.
Gemini-2.5-Flash achieves a 61.36\% average correlation at a cost of only \$0.131, whereas Claude-3.5-Sonnet requires \$2.774 (approximately 21 times higher) to reach a similar correlation of 61.65\%. This finding indicates that cost and performance are not strictly linearly coupled in HOBRE evaluation.

\begin{tcolorbox}[colback=gray!5,
                  colframe=black,
                  arc=0.8mm, auto outer arc,
                  boxrule=1pt,
                  boxsep=-2pt
                 ]
\textbf{Answering RQ2:} The configuration of backbone LLMs, prompting strategy, and decoding temperature exhibit significant and non-linear impacts on correlation and cost. The optimal prompting strategy (especially FSL and CoT) shows a performance inversion across LLMs'scale, while lower temperatures generally benefit the correlation of large-scale LLMs.
\end{tcolorbox}

\begin{table}[t]
    \caption{Cross-Task Kendall’s $\tau$ Correlation Analysis of Correlation and Cost Rankings for Judge Configurations.}
    \vspace{-1.3ex}
    \centering
    \renewcommand{\arraystretch}{1}
    \setlength{\tabcolsep}{1.1mm}
    \scalebox{0.89}{
    \begin{threeparttable}
    \begin{tabular}{cC{1.3cm}C{1.3cm}C{1.3cm}C{1.3cm}}
        \toprule
            \multirow{2}{*}{\textbf{Task-Task}} & \multicolumn{4}{c}{\textbf{Meta-Metrics}} \\
        \cmidrule(lr){2-5}
            & $\tau$ (\%) & $r_s$ (\%) & $r_p$ (\%) & Cost (\$)  \\
        \midrule
            FNR - BCS & 61.39 & 61.49 & 60.59 & 94.83 \\
            FNR - DO  & 55.38 & 56.28 & 61.18 & 95.98 \\
            BCS - DO  & 52.56 & 55.95 & 59.20 & 93.74  \\
        \bottomrule
    \end{tabular} 
    \end{threeparttable} }
    \vspace{-1ex}
    \label{tab:task}
\end{table}


\vspace{-2.4ex}
\subsection{RQ3: Optimal Configuration Variability} \label{sec:optimal}
Following the analysis of judge configurations in RQ2, this section further investigates whether a "one-size-fits-all" configuration exists that can consistently maintain high correlation and low cost across different task types and input samples.

\noindent\textbf{Task-level Configuration Variability.} As previously discussed, for each task, we construct 81 judge configurations by combining 9 backbone LLMs $\mathcal{M}$, 3 prompting strategies $\mathcal{P}$, and 3 decoding temperatures $\mathcal{T}$. A single configuration for task $i$ is defined as:
$$C_{i}(m, p, t) \rightarrow (\tau, r_s, r_p, \text{Cost})$$
where $m \in \mathcal{M}, p \in \mathcal{P}, t \in \mathcal{T}$. We calculate the Kendall’s $\tau$ correlation of all configurations across the three tasks in terms of the four metrics, and the results are shown in Table~\ref{tab:task}.

The rank correlations for the three performance metrics ($\tau, r_s, r_p$) across tasks are only at a moderate level. For example, the Kendall’s $\tau$ between BCS and DO is only 52.56\%. This "rank flipping" phenomenon proves that a configuration performing excellently in one task may fail to maintain its superiority in another, reflecting inherent disparities in the capability requirements of different HOBRE tasks. In contrast, the rank correlation for Cost across tasks is extremely high, indicating that costs are minimally affected by the task type. Therefore, when selecting configurations for different tasks, one can prioritize task characteristics to predict evaluation effectiveness, while treating cost as a relatively independent dimension.

\begin{table*}[t]
    \caption{Correlation and Cost of \method routing configuration compared to other configuration strategies and fine-tuned judge model ($\textcolor{customred}{\uparrow}$ represents the correlation improvement and $\textcolor{customgreen}{\times}$ represents the cost ratio of \method relative to each baseline) (All results are averaged over 5-fold cross-validation in \bench).}
    \vspace{-2.5ex}
    \centering
    \renewcommand{\arraystretch}{0.93}
    \setlength{\tabcolsep}{1.1mm}
    \scalebox{0.87}{
    \begin{threeparttable}
    \begin{tabular}{lccC{1.5cm}C{1.5cm}C{1.5cm}C{1.5cm}C{1.5cm}C{1.5cm}C{1.5cm}C{1.5cm}C{1.5cm}}
        \toprule
             \multicolumn{3}{c}{\textbf{Configuration}} & \multicolumn{3}{c}{\textbf{Function Name Recovery (FNR)}} & \multicolumn{3}{c}{\textbf{Binary Code Summarization (BCS)}} & \multicolumn{3}{c}{\textbf{Decompilation Optimization (DO)}}   \\
        \cmidrule(lr){1-3} \cmidrule(lr){4-6}  \cmidrule(lr){7-9}  \cmidrule(lr){10-12}  
            Models & Prompt & Temp. & \textbf{$\tau$ (\%)} & \textbf{$r_s$ (\%)}  & Cost (\$) & \textbf{$\tau$ (\%)} & \textbf{$r_s$ (\%)}  & Cost (\$)  & \textbf{$\tau$ (\%)} & \textbf{$r_s$ (\%)}  & Cost (\$)  \\
        \midrule
        \multicolumn{3}{l}{\textbf{Random Configuration}} 
        & $54.40_{\textcolor{customred}{(\uparrow35.1\%)}}$
        & $63.55_{\textcolor{customred}{(\uparrow30.2\%)}}$ 
        & $0.216_{\textcolor{customgreen}{(0.14\times)}}$ 
        & $55.44_{\textcolor{customred}{(\uparrow30.7\%)}}$ 
        & $65.09_{\textcolor{customred}{(\uparrow25.6\%)}}$ 
        & $0.207_{\textcolor{customgreen}{(0.15\times)}}$ 
        & $38.20_{\textcolor{customred}{(\uparrow65.5\%)}}$ 
        & $43.76_{\textcolor{customred}{(\uparrow62.2\%)}}$ 
        & $0.278_{\textcolor{customgreen}{(0.25\times)}}$ \\
        \cmidrule(lr){1-12}                    
            \multicolumn{3}{l}{\textbf{Static Configuration}} \\
            Claude-3.5-Sonnet & ZSL & 0.1 
            & $\textbf{70.33}_{\textcolor{customred}{(\uparrow4.5\%)}}$ 
            & $\textbf{78.99}_{\textcolor{customred}{(\uparrow4.8\%)}}$ 
            & $0.467_{\textcolor{customgreen}{(0.07\times)}}$ 
            & $60.17_{\textcolor{customred}{(\uparrow20.4\%)}}$ 
            & $69.16_{\textcolor{customred}{(\uparrow18.2\%)}}$ 
            & $0.493_{\textcolor{customgreen}{(0.06\times)}}$ 
            & $57.42_{\textcolor{customred}{(\uparrow10.1\%)}}$ 
            & $64.16_{\textcolor{customred}{(\uparrow10.6\%)}}$ 
            & $0.705_{\textcolor{customgreen}{(0.10\times)}}$  \\
            Gemini-2.5-Flash  & FSL & 0.5 
            & $68.06_{\textcolor{customred}{(\uparrow8.0\%)}}$ 
            & $77.79_{\textcolor{customred}{(\uparrow6.4\%)}}$ 
            & $0.071_{\textcolor{customgreen}{(0.44\times)}}$ 
            & $\textbf{68.10}_{\textcolor{customred}{(\uparrow6.4\%)}}$ 
            & $\textbf{77.61}_{\textcolor{customred}{(\uparrow5.3\%)}}$ 
            & $0.052_{\textcolor{customgreen}{(0.61\times)}}$ 
            & $50.67_{\textcolor{customred}{(\uparrow24.7\%)}}$ 
            & $57.88_{\textcolor{customred}{(\uparrow22.6\%)}}$ 
            & $0.110_{\textcolor{customgreen}{(0.63\times)}}$ \\
            DeepSeek-V3.2  & FSL & 0.1 
            & $65.64_{\textcolor{customred}{(\uparrow12.0\%)}}$ 
            & $74.41_{\textcolor{customred}{(\uparrow11.2\%)}}$ 
            & $0.053_{\textcolor{customgreen}{(0.58\times)}}$ 
            & $66.62_{\textcolor{customred}{(\uparrow8.8\%)}}$ 
            & $75.66_{\textcolor{customred}{(\uparrow8.1\%)}}$ 
            & $0.039_{\textcolor{customgreen}{(0.81\times)}}$ 
            & $\textbf{59.75}_{\textcolor{customred}{(\uparrow5.8\%)}}$ 
            & $\textbf{67.39}_{\textcolor{customred}{(\uparrow5.3\%)}}$ 
            & $0.082_{\textcolor{customgreen}{(0.84\times)}}$ \\
        \cmidrule(lr){1-12}                    
            \multicolumn{2}{l}{\textbf{Fine-tuned Judge Model}} \\
            UniXcoder &  &  
            & $53.89_{\textcolor{customred}{(\uparrow36.4\%)}}$ 
            & $68.05_{\textcolor{customred}{(\uparrow21.6\%)}}$ 
            & - 
            & $45.22_{\textcolor{customred}{(\uparrow60.2\%)}}$ 
            & $58.51_{\textcolor{customred}{(\uparrow39.7\%)}}$ 
            & -
            & $38.70_{\textcolor{customred}{(\uparrow63.3\%)}}$ 
            & $49.76_{\textcolor{customred}{(\uparrow42.6\%)}}$ 
            & - \\
            Qwen2.5-Coder-7B  &  &  
            & $69.27_{\textcolor{customred}{(\uparrow6.1\%)}}$ 
            & $77.75_{\textcolor{customred}{(\uparrow6.4\%)}}$ 
            & - 
            & $64.11_{\textcolor{customred}{(\uparrow13.0\%)}}$ 
            & $73.47_{\textcolor{customred}{(\uparrow11.3\%)}}$ 
            & -
            & $51.36_{\textcolor{customred}{(\uparrow23.1\%)}}$ 
            & $56.56_{\textcolor{customred}{(\uparrow25.5\%)}}$ 
            & - \\
        \cmidrule(lr){1-12}
            \rowcolor{gray!18}  \multicolumn{3}{l}{\textbf{\method Routing Configuration}} & \textbf{73.51} & \textbf{82.74} & 0.031 & \textbf{72.45} & \textbf{81.75} & 0.032 & \textbf{63.21} & \textbf{70.97} & 0.069 \\
        \cmidrule(lr){1-12}
            \multicolumn{3}{l}{\textbf{Oracle Configuration}} 
            & $92.72$ 
            & $95.08$ 
            & $0.019$ 
            & $98.40$ 
            & $99.04$ 
            & $0.014$ 
            & $90.29$ 
            & $92.18$ 
            & $0.040$ \\
        \bottomrule
    \end{tabular} 
    \end{threeparttable} } 
    \vspace{-2.5ex}
    \label{tab:binjudge}
\end{table*}

\noindent\textbf{Sample-level Configuration Variability.} To further verify whether the optimal configuration varies across samples, we compare the performance of three configuration selection strategies:

\setlength{\leftmargini}{9pt}
\begin{itemize}
\item \textbf{Random Configuration:} A configuration is randomly selected from the 81 groups for each sample.

\item \textbf{Static Configuration:} A single fixed configuration is used for all samples. We select the one with the highest overall correlation for each task as a representative.

\item \textbf{Oracle Configuration:} For each sample, we select the configuration that minimizes the squared error between its score and the human expert's rating (if multiple configurations achieve the same error, the one with the lowest cost is chosen).
This represents the upper bound of performance under an “ideal scenario”.
\end{itemize}

The results in Table~\ref{tab:optimal} show that Random Configuration is significantly worse than Static Configuration, indicating that improper selection severely degrades evaluation quality. Moreover, Static Configuration is far below Oracle Configuration, with a significant "Oracle Gap" existing across all three tasks. This quantitatively proves that using a single static configuration for all samples cannot reach the theoretical upper bound of evaluation performance, as any single configuration faces a clear "capability ceiling" when dealing with heterogeneous samples.

\vspace{-1ex}
\begin{tcolorbox}[colback=gray!5,
                  colframe=black,
                  arc=0.8mm, auto outer arc,
                  boxrule=0.85pt,
                  boxsep=-2.1pt
                 ]
\textbf{Answering RQ3:} There is no "one-size-fits-all" judge configuration. Different tasks have distinct configuration preferences, and even within the same task, the performance of a fixed configuration (max 70.31\% $\tau$) is far below the 92.95\% theoretical upper bound of sample-level adaptive configurations.
\end{tcolorbox}

\vspace{-3ex}
\section{\NoCaseChange{\method} Router}\label{sec:binjudge}
\vspace{-0.5ex}
\subsection{Adaptive Configuration Routing}
\vspace{-0.5ex}
Empirical results indicate that the optimal judge configuration dynamically varies with task types and sample characteristics, revealing a substantial Oracle Gap between the "single static best" configuration and the theoretical upper bound. 
This underscores the high heterogeneity of HOBRE evaluation tasks, where a one-size-fits-all configuration fails to accommodate all tasks and samples. Motivated by this, we introduce \method, a lightweight routing mechanism designed for task- and sample-aware adaptive configuration selection.

\method takes the stripped pseudocode and the task type as input, and dynamically routes the optimal judge configuration for each sample. Specifically, it employs a pretrained UniXcoder-base encoder to encode the input pseudocode, while freezing all parameters during training so that it serves as a fixed feature extractor. Building upon this, a learnable task embedding layer maps different task types into vector representations, which are then concatenated with the code embeddings to equip the model with cross-task awareness. After feature fusion, a stacked multi-layer perceptron is used to predict the preference distribution over all judge configurations.

During training, the optimization objective of \method is to maximize the sample-level utility function $U$, aiming to prioritize high correlation while reducing cost. The utility function of configuration $c$ on sample $i$ is defined as:
$$\begin{aligned}
\scriptsize
U_{i, c} &= - (\text{MSE}_{i, c} + \lambda \cdot \text{Cost}_{i, c}) \\[-5pt]
&= - \left( (S_{i, c} - h_i)^2 + \lambda \cdot \frac{\ln(1 + 10^3 \cdot \text{Cost}_{i, c})}{\ln(1 + 10^3 \cdot \max_{k \in \mathcal{C}_{valid}} \text{Cost}_{i, k} + \epsilon)} \right)
\end{aligned}$$
where $S_{i,c}$ denotes the score assigned by configuration $c$ to sample $i$, $h_i$ denotes the corresponding human expert score, and $\text{Cost}_{i,c}$ is the actual API expenditure. For the cost term, we apply logarithmic scaling for normalization, and set $\lambda = 0.1$ to control the trade-off between correlation and cost.

We adopt distribution alignment for training, where the utility $U$ is first transformed into a target probability distribution $P_{target}$ using a Softmax function with a temperature $T = 0.5$. The model's output layer then yields a predicted distribution $Q_\theta$ via a Log-Softmax operation. The training loss is defined as the KL divergence between the target and the predicted distribution:
$$\small \mathcal{L} = \text{KL}(P_{target} \parallel Q_{\theta}) = \frac{1}{N} \sum_{i=1}^{N} \sum_{c \in \mathcal{C}} P_{target}(c|i) \log \frac{P_{target}(c|i)}{Q_{\theta}(c|i)}$$
This training strategy enables \method not only to learn the top-ranked configuration, but also to capture the relative utility ordering of all 81 configurations for a given sample, thereby substantially improving robustness under limited training data.

During inference, for a new sample, \method requires only a single forward pass to obtain the predicted probability distribution over all configurations, and selects the configuration with the highest probability as the final judge configuration.
Subsequently, we conduct experiments to answer the following research question:

\setlength{\leftmargini}{9pt}
\begin{itemize}
\item \textbf{RQ4: Can \method routing effectively improve correlation and maintain low cost?} 
We explore the performance of \method in terms of correlation and cost compared to static optimal configuration, random configuration, and oracle upper bound, as well as two supervised fine-tuning judge models.
\end{itemize}

\vspace{-2ex}
\subsection{RQ4: Effectiveness of Configuration Routing} \label{sec:judge}
Considering that human expert annotations are limited and that obtaining additional annotations is costly, we evaluate the effectiveness of \method using five-fold cross-validation on the \bench dataset. Experimental results, as detailed in Table~\ref{tab:binjudge}, demonstrate that \method consistently outperforms all Static Configurations across the three HOBRE tasks. Specifically, our routing mechanism improves Kendall's $\tau$ by ranges of 4.5\%--12.0\%, 6.4\%--20.4\%, and 5.8\%--24.7\%, respectively, effectively narrowing the oracle gap. Beyond superior correlation with human experts, \method exhibits remarkable economic efficiency. The average API expenditure across all tasks is significantly lower than that of the Random Configuration, representing only 0.14$\times$, 0.15$\times$, and 0.25$\times$ of its cost, respectively. When compared to the Static Configuration, \method maintains a cost ratio of only 0.07--0.58$\times$, 0.06--0.81$\times$, and 0.10--0.84$\times$. Even relative to the Oracle Configuration, its cost is only about 2--3 times higher. These results validate the feasibility and practicality of learning-based configuration routing for scalable and accurate HOBRE evaluation.

Meanwhile, we fine-tuned UniXcoder and a lightweight LLM, Qwen2.5-Coder-7B~\cite{hui2024qwen2}, to directly predict scores using the same five-fold cross-validation. Both significantly underperformed BinJudge. This indicates that, under the same annotation budget, a fine-tuned judge generalizes poorly, whereas \method leverages complementary prior knowledge from different LLMs and adapts to sample variations via lightweight routing.


\begin{tcolorbox}[colback=gray!5,
                  colframe=black,
                  arc=0.8mm, auto outer arc,
                  boxrule=0.85pt,
                  boxsep=-2.1pt
                 ]
\textbf{Answering RQ4:} \method effectively bridges the gap between static and oracle configurations by adaptively selecting judge configurations for individual samples, improving Kendall's $\tau$ by 4.5\%–24.7\% and reducing API costs to 0.06$\times$–0.84$\times$ of static best configurations.
\end{tcolorbox}


\vspace{-1.8ex}
\section{Discussion}
\subsection{Analysis of Evaluator Disagreements}
We analyze the causes of disagreements across three dimensions:

\noindent\textbf{LLM vs. Traditional Metrics.} Discrepancies stem from two primary causes: (1) low-quality or incomplete references reduce the reliability of reference-based metrics; (2) traditional metrics underestimate outputs that are semantically equivalent but lexically different from the reference.

\noindent\textbf{LLM vs. Human.} As detailed in \S\ref{sec:compare}, our manual inspection reveals three main patterns leading to LLM-Human disagreements (i.e., $\vert{}\text{HumanScore} - \text{avg. LLMScore}\vert{} \ge 2$): (1) lack of semantic anchors, (2) lack of contextual information, and (3) rewarding fluent-but-unsupported outputs.

\noindent\textbf{Human vs. Human.} We define this disagreement as samples with a score range $\ge 2$ among the three annotators (253 out of 1233 samples; 73/105/75 for FNR/BCS/DO). Manual inspection shows this disagreement mainly occurs on medium-quality outputs due to intrinsic subjectivity, while annotators are generally consistent on clearly good or clearly poor outputs, with no clear pattern tied to specific input features. Compared with LLM-Human disagreement (149 out of 1233 samples; 70/18/61 for FNR/BCS/DO), the overlap is only 23 samples (12/6/5 for FNR/BCS/DO). This low overlap indicates that the two are largely distinct phenomena. We therefore do not claim LLM judges can fully replace humans, but that they are a superior option to traditional metrics.




\vspace{-1.5ex}
\subsection{Threats to Validity}\label{sec:threats}
\vspace{-0.5ex}
\textbf{Internal Validity.} The primary threat to internal validity relates to the inherent subjectivity of human evaluation used to construct the \bench ``gold standard''. Human assessments of ``readability'' and ``usefulness'' can vary significantly among individuals. To mitigate this threat, three experts from our author team (each with over three years of binary reverse engineering experience) annotated the dataset using a rigorous three-stage protocol: initial calibration, independent scoring, and a consensus meeting for high-variance samples. To prevent potential author bias, we further invited an independent industrial expert to annotate a subset (50 samples per task); merging these independent annotations with our original three-expert scores yields evaluation trends that remain highly consistent with original overall trends.

\noindent\textbf{External Validity.} Threats to external validity primarily concern the generalizability of our findings. To ensure a diverse and representative dataset, \bench encompasses 51 real-world GNU projects compiled across 6 target architectures and 4 optimization levels. We also evaluated responses generated by 8 distinct models, ranging from binary-specific deep learning models to ultra-large-scale general-purpose LLMs. This benchmark collectively spans multiple major HOBRE tasks: function name recovery, binary code summarization, and decompilation optimization. 

\noindent\textbf{Construct Validity.} Threats to construct validity concern whether our experimental design and evaluation metrics accurately capture the phenomena under investigation. To measure the alignment between automated approaches and human judgments, we adopted Kendall's $\tau$ rank correlation coefficient, Spearman's $r_s$ rank correlation coefficient, and Pearson's $r_p$ correlation coefficient. Following established methodologies in recent software engineering literature~\cite{zhou2025se, zeng2025evaluating}, the rank-based metrics ($\tau$ and $r_s$) are particularly well-suited for evaluating our ordinal Likert-scale data, while $r_p$ serves as a supplementary measure of linear correlation.

\vspace{-1.5ex}
\section{Conclusion and Future Work}\label{sec:conclusion}
\vspace{-0.5ex}
This paper presents the first systematic investigation into the LLM-as-a-Judge paradigm for Human-Oriented Binary Reverse Engineering (HOBRE) tasks. To overcome the inherent limitations of traditional match-based and test-based metrics, we introduced \bench, a comprehensive, expert-annotated, reference-free evaluation benchmark. Our extensive empirical study revealed that LLM evaluators significantly outperform traditional metrics in approximating human judgment, effectively capturing deep semantic equivalence rather than relying on superficial lexical overlaps. Furthermore, recognizing that no single judge configuration is optimal across all scenarios, we proposed \method, a lightweight, adaptive routing mechanism. By dynamically selecting the most suitable LLM configuration for each individual sample, \method substantially improves evaluation correlation while drastically reducing API costs. Overall, this work provides the binary reverse engineering community with a scalable, high-fidelity, and cost-effective framework for automated artifact evaluation.

\vspace{-1ex}
\section*{Data Availability Statement}
To support the reproducibility and replicability of the research, we release \bench and \method at \url{https://doi.org/10.5281/zenodo.19247812}. The released resources comprise all source code, datasets, and a subset of experimental results.

\begin{acks}
This research / project is supported by the National Key Research and Development Program of China under Grant 2025QY2840, and the Natural Science Foundation of China under Grant U20B2047, 62072421, 62002334, 62102386, and 62121002. This research / project is also supported by the National Research Foundation, under its Investigatorship Grant (NRF-NRFI08-2022-0002). Any opinions, findings and conclusions or recommendations expressed in this material are those of the author(s) and do not reflect the views of National Research Foundation, Singapore.
\end{acks}

\balance
\bibliographystyle{ACM-Reference-Format}
\bibliography{main}


\end{document}